\documentclass[prd,twocolumn,nofootinbib,showpacs,amsmath,amssymb,floatfix,eqsecnum]{revtex4-1}

\usepackage{graphicx,color,framed}
\usepackage{hyperref}
\usepackage{times}
\usepackage{enumerate}
\usepackage{lipsum}
\usepackage{slashed}

\hypersetup{
    colorlinks=true, %set true if you want colored links
    linktoc=all,     %set to all if you want both sections and subsections linked
    linkcolor=blue,  %choose some color if you want links to stand out
}

\def \beq {\begin{equation}}
\def \eeq {\end{equation}}
\def \beqa {\begin{eqnarray}}
\def \eeqa {\end{eqnarray}}
\def \bseq {\begin{subequations}}
\def \eseq {\end{subequations}}
\newcommand \dg {\dagger}

\newcommand \al {\alpha}

\newcommand \ran {\rangle}
\newcommand \lan {\langle}
\newcommand \un {\underline}
\newcommand \ep {\epsilon}
\newcommand \lam {\lambda}
\newcommand \pd {\partial}

\newcommand \mb {\mathbf}

\newcommand \nnb {\nonumber}

\newcommand \vphi {\varphi}

\newcommand \ov {\overline}
\newcommand \tomega {\tilde{\omega}}
\newcommand \td {\tilde}

\begin{document}

\title{Hall viscosity and geometric response in the Chern-Simons matrix model of the Laughlin states}

\author{Matthew F. Lapa}
\email[email address: ]{lapa2@illinois.edu}
\affiliation{Department of Physics and Institute for Condensed Matter Theory, University of Illinois at Urbana-Champaign, Urbana, IL, 61801-3080}
\author{Taylor L. Hughes}
\affiliation{Department of Physics and Institute for Condensed Matter Theory, University of Illinois at Urbana-Champaign, Urbana, IL, 61801-3080}

%\date{\today}

\begin{abstract}

We study geometric aspects of the Laughlin fractional quantum Hall (FQH) states using a description of these states in terms of
a matrix quantum mechanics model known as the Chern-Simons matrix model (CSMM). This model was proposed by 
Polychronakos as a regularization of the noncommutative Chern-Simons theory description of the Laughlin states proposed 
earlier by Susskind. Both models can be understood as describing the electrons in a FQH state as forming a noncommutative 
fluid, i.e., a fluid occupying a noncommutative space. Here we revisit the CSMM in light of recent work on geometric response in 
the FQH effect, with the goal of determining whether the CSMM captures this aspect of the physics of the Laughlin states. 
For this model we compute the Hall viscosity, Hall conductance in a non-uniform electric field, and the Hall viscosity in the 
presence of anisotropy (or intrinsic geometry). 
Our calculations show that the CSMM captures the guiding center contribution to the known
values of these quantities in the Laughlin states, but lacks the Landau orbit contribution. The interesting correlations in a  
Laughlin state are contained entirely in the guiding center part of the state/wave function, and so we conclude that the CSMM 
accurately describes the most important aspects of the physics of the Laughlin FQH states, including the Hall viscosity and
other geometric properties of these states which are of current interest.

\end{abstract}


\maketitle

%\textbf{MFL: Maybe change title to ``Quantum Hall physics in the Chern-Simons matrix model: Hall viscosity, 
%anisotropy, and Hall conductance in a non-uniform field".}

\section{Introduction}

In the past few years there has been tremendous progress in the understanding of the geometric properties of 
quantum Hall states. The role of geometry in the quantum Hall effect first came to prominence in early work on 
Hall viscosity~\cite{ASZ,levay,avron1998odd} 
(sometimes called \emph{odd} viscosity), and there has been much work on Hall viscosity since 
then~\cite{TV1,read2009,TV2,haldane2009,haldane2011,read-rezayi,HLF2011,hoyos-son,bradlyn2012,park-haldane}. 
Recent work on geometry in the quantum Hall effect has gone in two separate directions. First, there is the study of 
the properties of quantum Hall states on curved spatial manifolds 
(Riemann surfaces)~\cite{AG,cho2014,ferrari-klevtsov,BR1,BR2,CLW,framing,KN}. Second, there is the
study of intrinsic geometry and anisotropy in quantum Hall states on 
flat space~\cite{haldane2009,haldane2011,park-haldane,YCF-nematic,haldane-anisotropic}.
In the past year a very interesting
new theory of quantum Hall states has been proposed, known as the \emph{bi-metric} theory, and this theory promises
to unify the two separate directions of research on geometry in the quantum Hall effect~\cite{GGB,gromov-son}.

In a separate line of development, Susskind proposed in 2001 that Laughlin fractional quantum Hall (FQH) states
could be described by \emph{noncommutative Chern-Simons} (NCCS) theory~\cite{susskind}. 
This is a deformation of ordinary Chern-Simons theory in which the theory is formulated on a noncommutative analog of 
the flat space $\mathbb{R}^2$ consisting of ``coordinates" $\hat{x}^1$ and $\hat{x}^2$ obeying a nontrivial 
commutation relation 
\beq
	[\hat{x}^1,\hat{x}^2]=i\theta\ .
\eeq
Here $\theta$ is a real parameter with dimensions of length squared that characterizes the degree of 
noncommutativity of the theory. The original motivation for this proposal was Susskind's observation that the gauge
symmetry of NCCS theory provides a discretization of the symmetry under area-preserving diffeomorphisms that is present
in a description of a FQH state as a charged fluid in a magnetic field. In particular, this discretization was argued to capture
the ``granularity" of a fluid constructed from electrons, and in the NCCS theory description each electron is associated with a
fundamental area of size $2\pi|\theta|$. In addition, in the NCCS theory a quantization rule~\cite{PN} enforces 
\beq
	\theta= \ell_B^2 m\ ,\ m\in\mathbb{Z}\ ,
\eeq
where $\ell_B$ is the magnetic length, and so one finds (for $m>0$) that the fluid described by the NCCS theory has a number 
density $\rho_0=\frac{1}{2\pi\ell_B^2 m}$, exactly as in the $\nu=\frac{1}{m}$ Laughlin state.

Susskind's original proposal has been followed up by many 
authors~\cite{P1,P3,MP,HVR,karabali-sakita1,karabali-sakita2,hansson2001,fradkin-NCCS,hansson2003,cappelli2005}. 
Of all of these subsequent works, the
work of Polychronakos is particularly important for this article. In Ref.~\onlinecite{P1}, Polychronakos proposed a
regularization of the NCCS theory which is known as the \emph{Chern-Simons matrix model} (CSMM).
This is a particular regularization of the NCCS theory which can be viewed as a quantum mechanics model in which the
degrees of freedom are $N\times N$ matrices
(we discuss the model in detail and make this statement precise below). The quantum ground state
of the CSMM having $\theta=\ell_B^2 m$ ($m>0$) is known to describe a uniform droplet of ``noncommutative fluid" with 
a density and area matching that of the $\nu=\frac{1}{m}$ Laughlin state. Polychronakos has also demonstrated that
excitations in this model carry the appropriate fractional charge of the quasihole excitations in the Laughlin state.

Despite the successes in describing the basic properties of the Laughlin FQH states using NCCS theory and the CSMM, 
there have not been any attempts to study \emph{geometric} properties of the Laughlin states in the context of these
noncommutative models. Therefore, our goal in this article is to answer the following question: \emph{does the CSMM 
accurately describe the geometric properties of the Laughlin states?} 

The particular geometric properties that we are
concerned with are the Hall viscosity, the Hall conductance in a non-uniform electric field, and the
Hall viscosity in the presence of anisotropy (or intrinsic geometry). 
We compute all of these quantities in the CSMM and we find that the
results in the CSMM contain only the \emph{guiding center} contribution to the known values for these quantities in the
Laughlin states. For example, the full Hall viscosity coefficient
for the Laughlin $\nu=\frac{1}{m}$ state is given by~\cite{read2009}
\beq
	\eta_{tot}= \frac{\hbar \rho_0 m}{4}\ ,
\eeq 
while for the CSMM with $\theta=\ell_B^2 m$ we find\footnote{In the literature the quantity $\frac{m-1}{2}$ is referred
to either as the \emph{anisospin} (Refs.~\onlinecite{GGB,gromov-son}) or minus the \emph{guiding center spin}
(Refs.~\onlinecite{haldane2009,haldane2011,park-haldane}) of the $\nu=\frac{1}{m}$ Laughlin state.}
 (after regularization) 
\beq
	\eta_{\text{{\tiny{CSMM}}},reg}= \frac{1}{2}\hbar \rho_0 \left(\frac{m-1}{2}\right)\ ,
\eeq
which is exactly the (regularized) \emph{guiding center Hall viscosity} of the $\nu=\frac{1}{m}$ 
Laughlin state~\cite{haldane2009,haldane2011,park-haldane}. The need for regularization of the
guiding center part of the Hall viscosity has been discussed in Refs.~\onlinecite{haldane2009,haldane2011,park-haldane}. 
In this paper we also give a fluid
interpretation of this regularization in the context of the CSMM.

Based on our calculations we conclude quite generally that the CSMM and NCCS theory descriptions of the Laughlin 
FQH states capture the guiding center contribution to the geometric properties of these states, but lack the Landau
orbit contribution. We argue that this is not surprising since in the fluid interpretation of the CSMM and
NCCS theories, the cyclotron frequency $\omega_c$ is sent to infinity by sending the mass of the particles in the fluid
to zero. This is analogous to a projection into a Landau level (which freezes out the Landau orbit degrees of freedom), and so it 
makes sense that only the guiding center contribution remains.
The Landau orbit contribution is often considered to be less important since the interesting correlations in a Laughlin
state are contained entirely in the guiding center part of the state/wave function. Therefore we find that the CSMM
description is able to capture the most important contributions to the geometric properties of the Laughlin 
states. We expect that our results will rekindle interest in noncommutative
models of the FQH effect, as these models clearly have a role to play in the investigation of geometric properties of 
FQH states.

This paper is organized as follows. In Sec.~\ref{sec:GC-hall-visc} we review the notion of Hall viscosity. 
In Secs.~\ref{sec:NCCS} and \ref{sec:CSMM} we give a comprehensive review of the NCCS theory and CSMM, the
fluid interpretation of these models, and their relation to the Laughlin states. In Sec.~\ref{sec:CSMM-viscosity} we
compute the Hall viscosity in the CSMM. In Sec.~\ref{sec:hall-conductance} we compute the Hall conductance of 
the CSMM in a non-uniform electric field. In Sec.~\ref{sec:reg} we give a fluid interpretation of the regularization of the
guiding center part of the Hall viscosity in which one subtracts the extensive contribution to this quantity. Finally, in 
Sec.~\ref{sec:intrinsic} we present a modified version of the CSMM incorporating anisotropy, 
and we compute the Hall viscosity for the modified model. Sec.~\ref{sec:conclusion} presents our 
conclusions. The paper also includes two appendices. 
In Appendix~\ref{app:generators} we review the form of the quantum generators of the 
action of the group $U(N)$ on the fields of the CSMM, as this information is necessary for the quantization of this model which 
we review in Sec.~\ref{sec:CSMM}. In Appendix~\ref{app:Kubo} we present the details of the calculation of the Hall
viscosity of the CSMM (which is presented in Sec.~\ref{sec:CSMM-viscosity} of the main text), 
which involves a Kubo formula approach inspired by Ref.~\onlinecite{bradlyn2012}.

\section{Review of Hall viscosity}
\label{sec:GC-hall-visc}

In this section we review the concept of Hall viscosity following the derivation and point of view in 
Ref.~\onlinecite{park-haldane}. We also emphasize, again following Ref.~\onlinecite{park-haldane}, 
the separation of the Hall viscosity tensor into two parts: the \emph{Landau orbit} contribution and the \emph{guiding center}
contribution. Finally, we review the form of both parts of the Hall viscosity tensor for typical FQH trial states including the 
Laughlin states. The example of the Laughlin states is of particular interest for the rest of the paper when we 
compare to the results obtained in the CSMM, which has been argued to describe the physics of the Laughlin states.

\subsection{Hall viscosity calculation}

The Hall viscosity can be computed by studying the response of a FQH state to time-dependent area-preserving 
deformations (APDs). Before we review the calculation of the Hall viscosity, we briefly recall 
the setup of the quantum Hall problem. We consider $N$ electrons on the plane, each with a charge 
$-e<0$, in the presence of a constant background magnetic field of strength $B >0$ and pointing in the positive $z$ direction.
Let $\mb{r}_j$ be the position 
coordinates of the $N$ electrons, where $j=1,\dots,N,$ is a particle label. We write $r^a_j$ with $a=1,2,$ for the 
two components of the vector $\mb{r}_j$ (i.e., $a=1,2$, labels the two directions of space).
In this situation the electron coordinate operators $r^a_j$ break up into two parts as
\beq
	r^a_j = R^a_j + \tilde{R}^a_j\ ,
\eeq
where $R^a_j$ are known as the guiding center coordinates, and $\tilde{R}^a_j$ are the Landau orbit 
coordinates. These coordinates obey the commutation relations 
\begin{subequations}
\label{eq:coordinate-comm-rels}
\beqa
	\left[R^a_j, R^b_k\right] &=& i\ell_B^2 \ep^{ab}\delta_{jk} \\
	\left[\tilde{R}^a_j, \tilde{R}^b_k\right] &=& -i\ell_B^2 \ep^{ab}\delta_{jk} \\
	\left[R^a_j, \tilde{R}^b_k\right] &=& 0\ ,
\eeqa
\end{subequations}
where $\ell_B^2= \frac{\hbar}{eB}$ is the square of the magnetic length $\ell_B$.

The Hall viscosity is defined as the response of the system (more precisely, the ground state) to time-dependent
APDs of the electron coordinates. These APDs are generated by Hermitian operators $\mathsf{\Lambda}^{ab}$
which are a linear combination of guiding center and Landau orbit parts,
\beq
	\mathsf{\Lambda}^{ab}= \Lambda^{ab}-\td{\Lambda}^{ab}\ .
\eeq
The operators $\Lambda^{ab}$ generate APDs of the guiding center coordinates and have the form
\beq
	\Lambda^{ab}= \frac{1}{4\ell_B^2}\sum_{j=1}^N \left\{R^a_j,R^b_j\right\}\ ,
\eeq
where $\{\cdot,\cdot\}$ denotes an anti-commutator, while $\td{\Lambda}^{ab}$ generates APDs of the Landau orbit 
coordinates, and $\td{\Lambda}^{ab}$ is defined like $\Lambda^{ab}$ but with
the guiding center coordinates $R^a_j$ replaced by the Landau orbit coordinates $\td{R}^a_j$.
One can show that these generators obey the Lie algebras
\begin{subequations}
\label{eq:APD-algebra}
\begin{align}
	[\Lambda^{ab},\Lambda^{cd}] &= \frac{i}{2}\left(\ep^{bc}\Lambda^{ad} + \ep^{bd}\Lambda^{ac} + \ep^{ac}\Lambda^{bd}+\ep^{ad}\Lambda^{bc}  \right) \label{eq:APD-algebra-gc} \\
	\left[\td{\Lambda}^{ab},\td{\Lambda}^{cd}\right] &=  -\frac{i}{2}\left(\ep^{bc}\td{\Lambda}^{ad} + \ep^{bd}\td{\Lambda}^{ac} + \ep^{ac}\td{\Lambda}^{bd}+\ep^{ad}\td{\Lambda}^{bc}  \right) \ .
\end{align}
\end{subequations}
In addition, it is clear that $[\Lambda^{ab},\td{\Lambda}^{cd}]=0$.
The generators $\Lambda^{ab}$ (and also $\td{\Lambda}^{ab}$) can be expressed in terms of the generators of the Lie 
algebra of the group $SU(1,1)$, and we will
use this fact later\footnote{Physicists can read about the group $SU(1,1)$ in Ref.~\onlinecite{perelomov}, for example}.

Finite (as opposed to infinitesimal) APDs of the electron coordinates are implemented by conjugation by 
the unitary operators\footnote{Here, 
and in the rest of the article, we use a summation convention in which we sum over all indices which are repeated once as
a subscript and once as a superscript. All other summations will be indicated explicitly.}
\beq
	U(\al)= e^{i\al_{ab}\mathsf{\Lambda}^{ab}}\ ,
\eeq 
where $\al_{ab}$ is a constant, symmetric tensor with unit determinant (thus, the APDs are spatially uniform since $\al_{ab}$
does not depend on the spatial coordinates). For example, acting on the electron
coordinates gives
\beq
	U(\al)r^a_j U(\al)^{\dg}= r^a_j + \ep^{ab}\al_{bc}r^c_j + \dots\ ,
\eeq
where the ellipses denote higher order terms in $\al_{ab}$.  

The APDs that we have been considering so far are closely related to strains in continuum mechanics. Suppose
the vector $\mb{r}$ is the location of a point in a solid or fluid before a deformation, and $\mb{r}'(\mb{r})$ is the location of 
that same point after the deformation. Then for small deformations the \emph{strain tensor} $u_{ab}$ is defined in terms of 
the displacement vector $\mb{u}(\mb{r})= \mb{r}'(\mb{r})- \mb{r}$ as
\beq
	u_{ab}= \frac{1}{2}\left(\frac{\pd u_a}{\pd r^b} + \frac{\pd u_b}{\pd r^a} \right)\ ,
\eeq
where $u^a$ are the components of $\mb{u}(\mb{r})$ and $u_a= \delta_{ab}u^b$.
If we consider small APDs in the quantum Hall problem (i.e., we work to linear order in $\al_{ab}$), then we find a strain
tensor 
\beq
	u_{ab} = \frac{1}{2}\left( \delta_{ac}\ep^{cd}\al_{db} + (a\leftrightarrow b) \right)\ .
\eeq
In particular, we find for the trace $\sum_{a=1}^2 u_{aa}= 0$, which means that the APDs are indeed area-preserving (the
trace of the strain tensor determines the change in the area of a small element of the fluid or solid at the
location $\mb{r}$). The strain
tensor is also spatially uniform since $\al_{ab}$ does not depend on the spatial coordinates $r^a$. Therefore, the
APDs that we have been considering can be understood as a special case of a strain in continuum mechanics, namely, a
spatially uniform and area-preserving strain. In what follows we sometimes use the terms APD and strain interchangeably
although, strictly speaking, the former is a special case of the latter.

Consider a FQH system described by a Hamiltonian $H_0$. 
Under a time-independent APD parametrized by $\al_{ab}$ the Hamiltonian is transformed to 
\beq
	H(\al)= U(\al)H_0 U(\al)^{\dg}\ .
\eeq
We can define the generalized force associated with this APD as
\beq
	F^{ab}= -\frac{\pd H(\al)}{\pd \al_{ab}}\Big|_{\al=0} = -i[\mathsf{\Lambda}^{ab},H_0]\ .
\eeq
If we subject the system to a time-dependent APD $\al_{ab}(t)$, then we can compute the expectation
value of $F^{ab}$ in the time-dependent ground state $|\psi(t)\ran$ in an expansion in time derivatives of 
$\al_{ab}(t)$. In fact, as was argued in Ref.~\onlinecite{bradlyn2012}, one should actually compute the expectation value of  
$U(\al(t))F^{ab}U(\al(t))^{\dg}$ instead of $F^{ab}$. We discuss this point in more detail in the context of our
Kubo formula calculation of the Hall viscosity for the CSMM in Appendix~\ref{app:Kubo}, but 
just mention here that this replacement corresponds to expressing the generalized force in terms of the coordinates of the
deformed system. 

We now compute the expectation value of $U(\al(t))F^{ab}U(\al(t))^{\dg}$ in an expansion
in time derivatives of $\al_{ab}(t)$ as 
\begin{align}
	\lan \psi(t)| U(\al(t))F^{ab}U(\al(t))^{\dg} |\psi(t)\ran &= \nnb \\
 \lan \psi_0| F^{ab}|\psi_0\ran &+ \Gamma^{abcd}\dot{\al}_{cd}(t) + \dots\ ,
\end{align}
where $|\psi_0\ran$ denotes the initial state of the system, the overdot on $\al_{cd}(t)$ denotes a time derivative,
 and $\Gamma^{abcd}$ is a four index tensor which is 
independent of the parameters $\al_{ab}(t)$ (in principle there could also be an elastic term which is proportional
to $\al_{ab}(t)$, but this term is not present for a fluid state). Park and Haldane then define the full Hall viscosity 
tensor $\eta^{abcd}_{tot}$ (with all indices raised) as 
\beq
	\eta^{abcd}_{tot}= \frac{\Gamma^{abcd}}{A}\ ,
\eeq
where $A$ denotes the area of the quantum Hall droplet (recall that we are working on the infinite plane, so we must
assume that the quantum Hall droplet occupies a finite area $A$). The 
intuition behind this definition is that $\eta^{abcd}_{tot}$ encodes the linear response of the ``generalized stress"
$\frac{U(\al(t))F^{ab}U(\al(t))^{\dg} }{A}$ to the ``rate of strain" encoded by the tensor $\dot{\al}_{cd}(t)$.
We also note here that for a droplet of quantum Hall fluid the area $A$ of the droplet can be expressed as 
$A= 2\pi\ell_B^2 N_{\phi}$, where $N_{\phi}$ is the number of fundamental flux quanta $\Phi_0= \frac{h}{e}$ piercing the 
droplet.

Using adiabatic perturbation theory, Park and Haldane showed that 
\begin{align}
	\eta^{abcd}_{tot} &= \frac{i\hbar}{A}\lan \psi_0|[\mathsf{\Lambda}^{ab},\mathsf{\Lambda}^{cd}]|\psi_0\ran \nnb \\
	&=  \frac{i\hbar}{A}\lan \psi_0|[\Lambda^{ab},\Lambda^{cd}]|\psi_0\ran+\frac{i\hbar}{A}\lan \psi_0|[\td{\Lambda}^{ab},\td{\Lambda}^{cd}]|\psi_0\ran \nnb \\
	&\equiv \eta^{abcd}_H + \td{\eta}^{abcd}_H\ .
\end{align}
Thus, the full Hall viscosity tensor breaks up into two parts: the guiding center Hall viscosity tensor 
$\eta^{abcd}_H$, and the Landau orbit Hall viscosity tensor $\td{\eta}^{abcd}_H$.

The expression for the full Hall viscosity tensor can be simplified further by using the algebra of APD generators 
from Eq.~\eqref{eq:APD-algebra} to find
\beq
	\eta^{abcd}_{tot} = \frac{1}{2}\left( \ep^{ac}\eta^{bd}_{tot} + \ep^{ad}\eta^{bc}_{tot} + (a\leftrightarrow b)\right)\ ,
\eeq
where the symmetric two-index tensor $\eta^{ab}_{tot}$ also breaks up into guiding center and Landau orbit parts as
\beq
	\eta^{ab}_{tot}= \eta^{ab}_H + \td{\eta}^{ab}_H
\eeq
with
\begin{subequations}
\beqa
	\eta^{ab}_H &=& -\frac{\hbar}{A}\lan \psi_0|\Lambda^{ab}|\psi_0\ran \\
	\td{\eta}^{ab}_H &=& \frac{\hbar}{A}\lan \psi_0|\td{\Lambda}^{ab}|\psi_0\ran\ .
\eeqa
\end{subequations}
In what follows we also refer to these two-index tensors as ``Hall viscosity tensors".
Ref.~\onlinecite{park-haldane} emphasized that the guiding center contribution $\eta^{ab}_H$ to $\eta^{ab}_{tot}$ 
has a physical interpretation in 
terms of the intrinsic electric dipole moment along the edge of a FQH state, and in fact must be proportional to the
symmetric tensor which determines this dipole moment in order to balance the force 
on a FQH edge in an inhomogeneous electric field (see also Ref.~\onlinecite{wiegmann2012} for a complementary discussion of 
this boundary dipole moment from a different point of view). We now review the form of the two parts of the 
Hall viscosity tensor for typical FQH trial states including the Laughlin states.

\subsection{Values in quantum Hall trial states}

In this section we consider the form of the guiding center and Landau orbit Hall viscosity tensors $\eta^{ab}_H$ and
$\td{\eta}^{ab}_H$ for typical FQH trial states including the Laughlin states. In the operator, or 
Heisenberg, approach (as 
opposed to the Schrodinger approach using wave functions) a state vector for a trial FQH state is constructed using ladder operators $b_j$ and $b^{\dg}_j$ defined in terms of the 
guiding center coordinates as
\beq
	b_j= \frac{1}{\ell_B \sqrt{2}}(R^1_j + iR^2_j)\ ,
\eeq
and also ladder operators $a_j$ and $a^{\dg}_j$ defined in terms of the Landau orbit coordinates as
\beq
	a_j= \frac{1}{\ell_B\sqrt{2}}(\td{R}^1_j - i\td{R}^2_j)\ .
\eeq
We define $|0\ran_a$ and $|0\ran_b$ to be the Fock vacuum states annihilated by the $a_j$ and $b_j$ operators, 
respectively. In terms of these, a typical FQH trial state in the $n^{th}$ Landau level has 
the form
\beq
	|\psi_0\ran= \left[\prod_{j=1}^N \frac{(a^{\dg}_j)^n}{\sqrt{n!}}\right] F(b^{\dg}_1,\dots,b^{\dg}_N)|0\ran_a \otimes |0\ran_b\ ,
\eeq
where $F(b^{\dg}_1,\dots,b^{\dg}_N)$ is a homogeneous polynomial of $N$ variables, and which is either symmetric (for 
bosons) or antisymmetric (for fermions) under exchange of any two variables. We use $\text{Deg}[F]$ to denote the total 
degree of the polynomial function $F$. Then if we scale all arguments of $F$ by a numerical factor $\lambda$, we have
\beq
	 F(\lambda b^{\dg}_1,\dots,\lambda b^{\dg}_N)= \lambda^{\text{Deg}[F]} F(b^{\dg}_1,\dots,b^{\dg}_N)\ .
\eeq
Let $\mathcal{N}_b= \sum_{j=1}^N b^{\dg}_j b_j$ be the total number operator for the $N$ guiding center ladder 
operators. Then the homogeneity property of $F$ implies that $|\psi_0\ran$ is an eigenvalue of $\mathcal{N}_b$ with 
eigenvalue $\text{Deg}[F]$.

To compute $\eta^{ab}_H$ for these trial FQH states we use a connection between the
APD generators and the generators of the group $SU(1,1)$ (see, for example, Ref.~\onlinecite{perelomov}). 
Define the operators
\begin{subequations}
\beqa
	K_0 &=& \frac{1}{2}\sum_{j=1}^N\left(b^{\dg}_j b_j + \frac{1}{2}\right) \\
	K_{+} &=& \frac{1}{2}\sum_{j=1}^N (b^{\dg}_j)^2 \\
	K_{-} &=& \frac{1}{2}\sum_{j=1}^N (b_j)^2 \ .
\eeqa
\end{subequations}
These operators obey the commutation relations of the Lie algebra of the group $SU(1,1)$,
\begin{subequations}
\beqa
	[K_0,K_{\pm}] &=& \pm K_{\pm} \\
	\left[K_{-},K_{+}\right] &=& 2K_0\ .
\eeqa
\end{subequations}
The Fock space of the oscillators $b_j$ forms a (reducible) 
representation of this algebra, and the generators $\Lambda^{ab}$ can be expressed in terms of the $SU(1,1)$ generators
as
\beqa
	\Lambda^{11} &=& K_0 + \frac{1}{2}K_{+} + \frac{1}{2}K_{-} \\
	\Lambda^{22} &=& K_0 - \frac{1}{2}K_{+} - \frac{1}{2}K_{-} 
\eeqa
and
\beq
	\Lambda^{12}=\Lambda^{21}= \frac{-i}{2}(K_{-}-K_{+})\ .
\eeq

It is clear that the state $|\psi_0\ran$ is an eigenstate of $K_0$ with eigenvalue
$\frac{1}{2}(\text{Deg}[F]+\frac{N}{2})$. It then follows that the expectation values $\lan \psi_0 |K_{\pm}| \psi_0\ran$
are equal to zero as $K_{\pm}|\psi_0\ran$ is orthogonal to $|\psi_0\ran$. Then, for the trial state
parametrized by the function $F,$ we have
\beq
	\lan \psi_0|\Lambda^{ab}|\psi_0\ran= \frac{1}{2}\left[\text{Deg}[F]+\frac{N}{2}\right]\delta^{ab}\ .
\eeq
A similar computation shows that for a trial state in the $n^{th}$ Landau level we have
\beq
	\lan \psi_0|\td{\Lambda}^{ab}|\psi_0\ran = \frac{1}{2}\left(nN+\frac{N}{2}\right)\delta^{ab}\ ,
\eeq
which follows since the product $\prod_{j=1}^N \frac{(a^{\dg}_j)^n}{\sqrt{n!}}$ is a homogeneous polynomial in the
$a^{\dg}_j$ of total degree $n N$.

For the case of the $\nu= \frac{1}{m}$ Laughlin state ($m$ a positive integer) we have
\beq
	F(b^{\dg}_1,\dots,b^{\dg}_N)= \prod_{j<k}(b^{\dg}_j - b^{\dg}_k)^{m}\ ,
\eeq
and so
\beq
	\text{Deg}[F]= \frac{1}{2}m N (N-1)\ .
\eeq
If we consider this Laughlin state in the lowest Landau level ($n=0$) then we find that
\begin{subequations}
\label{eq:strain-Laughlin}
\begin{align}
	\lan \psi_0|\Lambda^{ab}|\psi_0\ran &= \frac{1}{2}\left[\frac{1}{2}m N^2 + \left( \frac{1-m}{2} \right)N\right]\delta^{ab} \\
	\lan \psi_0|\td{\Lambda}^{ab}|\psi_0\ran &= \frac{N}{4}\delta^{ab}\ ,
\end{align}
\end{subequations}
and so
\beq
	\eta^{ab}_H= -\frac{\hbar}{A}\frac{1}{2}\left[\frac{1}{2}m N^2 + \left( \frac{1-m}{2} \right)N\right]\delta^{ab}\ ,
\eeq
while 
\beq
	\td{\eta}^{ab}_H= \frac{\hbar}{4}\frac{N}{A}\delta^{ab}\ .
\eeq
Both of these tensors are proportional to the identity matrix (in this rotation-invariant case), and it is convenient to denote the 
constants of proportionality by
\beq
	\eta_H= -\frac{\hbar}{A}\frac{1}{2}\left[\frac{1}{2}m N^2 + \left( \frac{1-m}{2} \right)N\right]
\eeq
and 
\beq
	\td{\eta}_H=  \frac{\hbar}{4}\frac{N}{A}
\eeq
so that we can simply write $\eta^{ab}_H= \eta_H \delta^{ab}$ and similarly for $\td{\eta}^{ab}_H$.

For a Laughlin FQH droplet with $\nu=\frac{1}{m},$ and consisting of a large number $N$ of particles, we have 
$A\approx 2\pi \ell_B^2 m N$.
Then, in its current form, the coefficient $\eta_H$ in the guiding center Hall viscosity tensor is the sum of an extensive 
(order $N$) term and an intensive (order $1$) term. Since $A$ itself is proportional to $N$, the extensive term in $\eta_H$ 
comes from the superextensive (order $N^2$) term in $\lan \psi_0|\Lambda^{ab}|\psi_0\ran$. 
This term is associated with a uniform rotational 
motion (in fact, it is just the orbital angular momentum) of the FQH fluid, and
so it has been argued that one should subtract this term when defining the guiding center Hall 
viscosity~\cite{haldane2011,park-haldane}. If we make this subtraction then we end up with the regularized quantities
\beqa
	\lan \psi_0|\Lambda^{ab}|\psi_0\ran_{reg} &=& \frac{1}{2}\left[\left( \frac{1-m}{2} \right)N\right]\delta^{ab} \\
	\eta_{H,reg} &=& -\frac{\hbar}{2}\left( \frac{1-m}{2} \right)\rho_0 \ ,
\eeqa
where $\rho_0= \frac{1}{2\pi \ell_B^2 m}= \frac{N}{A}$ is the density of the $\nu=\frac{1}{m}$ Laughlin FQH state at large 
$N$. We discuss the physical interpretation of this regularization scheme in the context of the CSMM
in Sec.~\ref{sec:reg}.

The Landau orbit contribution $\td{\eta}_H$ does not require regularization as it only consists of an intensive term. In terms
of the density $\rho_0$ of the Laughlin state this coefficient has the form
\beq
	\tilde{\eta}_H= \frac{\hbar\rho_0}{4}\ .
\eeq
Then the full Hall viscosity coefficient for the $\nu=\frac{1}{m}$ Laughlin state (in the lowest Landau level and 
after regularization of the guiding center part) is
\beq
	\eta_{tot}=\tilde{\eta}_H+\eta_{H,reg}= \frac{\hbar\rho_0 m}{4}\ ,
\eeq
as originally found by Read~\cite{read2009}. It is interesting to observe that since $\rho_0=\frac{1}{2\pi\ell_B^2 m}$, the
full Hall viscosity coefficient $\eta_{tot}$ actually does not depend on the filling fraction of the Laughlin state (i.e., it 
does not depend on $m$).

The coefficient $\frac{1-m}{2}$ appearing in $\eta_{H,reg}$ is what Haldane has termed the ``guiding center spin"
of a FQH state. This
coefficient has been denoted as ``$\ov{s}$" in Ref.~\onlinecite{haldane2011} and ``$s$" in Ref.~\onlinecite{park-haldane}.
It is also equal to minus the ``anisospin" defined in Refs.~\onlinecite{GGB,gromov-son}, and denoted there by 
$\varsigma$. We choose to adopt the notation of  Refs.~\onlinecite{GGB,gromov-son} and so we write
\beq
	\eta_{H,reg} = \frac{\hbar}{2}\varsigma\rho_0  \label{eq:GC-visc-reg}
\eeq
with $\varsigma= \frac{m-1}{2}$. We see that unlike the full Hall viscosity coefficient $\eta_{tot}$, the
guiding center contribution to the Hall viscosity has a clear dependence on $m$.
It follows that different Laughlin states cannot be distinguished by their full Hall viscosity $\eta_{tot}$, but
they \emph{can} be distinguished by their guiding center Hall viscosity $\eta_{H,reg}$ which, moreover, has been argued to be
connected to the physical property of intrinsic electric dipole moment at the edge of the FQH state~\cite{park-haldane}.

%
% In this form the formula for $\eta_{H,reg}$ closely resembles the formulas
%for Hall viscosity obtained in Refs.~\onlinecite{read2009,read-rezayi}, 
%although we remind the reader that here we only consider the 
%guiding center part of the Hall viscosity, whereas Refs.~\onlinecite{read2009,read-rezayi} considered the full Hall viscosity 
%including the Landau orbit contribution.
%
%The full Hall viscosity coefficient $\eta_{tot}$ can be obtained by adding to the guiding center Hall viscosity $\eta_{H,reg}$
%the Landau orbit Hall viscosity $\tilde{\eta}_H$  (see Ref.~\onlinecite{park-haldane} for further discussion of these two 
%contributions to the full Hall viscosity). For the lowest Landau level, the Landau
%orbit contribution to the Hall viscosity is~\cite{ASZ,park-haldane}
%\beq
%	\tilde{\eta}_H= \frac{\hbar\rho_0}{4}\ .
%\eeq
%Then the full Hall viscosity for the $\nu=\frac{1}{m}$ Laughlin state is
%\beq
%	\eta_{tot}=\tilde{\eta}_H+\eta_{H,reg}= \frac{\hbar\rho_0 m}{4}\ ,
%\eeq
%as originally found by Read~\cite{read2009}. It is interesting to observe that since $\rho_0=\frac{1}{2\pi\ell_B^2 m}$, the
%full Hall viscosity coefficient $\eta_{tot}$ actually does not depend on the filling fraction of the Laughlin state. 

\section{Noncommutative Chern-Simons theory}
\label{sec:NCCS}

In this section we review Susskind's noncommutative Chern-Simons (NCCS) theory description of the Laughlin FQH 
states~\cite{susskind}. This will pave the way for the discussion of the Chern-Simons matrix model in the next section,
as the Chern-Simons matrix model can be thought of as a particular regularization of the NCCS theory. To prepare the reader 
for this discussion in this section we first make a few remarks about the two different formulations (``operator" vs. 
``star product" formulations) of noncommutative field theory. We then present the NCCS theory in both formulations. Finally,
we discuss the NCCS theory in the limit of weak noncommutativity, and its connection with the dynamics of a fluid of 
charged particles in a magnetic field. From this connection one sees that the full NCCS theory should be understood as
describing a fluid of charged particles in a magnetic field on a noncommutative space. 
Our discussion of noncommutative field
theory closely follows that in Refs.~\onlinecite{DN,harvey,szabo}. For the fluid picture of the NCCS theory we follow
Refs.~\onlinecite{susskind,P2}. Readers who are already familiar with noncommutative field theory and the NCCS theory
may want to skip this section.

\subsection{Two formulations of noncommutative field theory}

Consider a classical field theory in $2+1$ dimensions in which the two-dimensional space is taken to be $\mathbb{R}^2$, and
let $\mb{x}= (x^1,x^2)$ denote the spatial coordinates. We denote a general field in this theory as
$\Phi(t,\mb{x})$. In such a field theory the fields $\Phi(t,\mb{x})$ at a fixed time $t$ are elements of the ordinary algebra of 
functions on $\mathbb{R}^2$ (the commutative algebra generated by pointwise addition and multiplication of functions of
$\mb{x}$). The noncommutative deformation of the this theory that we consider consists of replacing
the ordinary space $\mathbb{R}^2$ with a ``noncommutative plane" whose two spatial coordinates do not commute with
each other. The time direction will always be commutative in this article, i.e., we consider
theories in two noncommutative spatial dimensions and one commutative (or ordinary) time direction. 

In the noncommutative deformation of the classical field theory, the fields (again at a fixed time $t$) instead take values in the
algebra $\mathbb{R}^2_{\theta}$ which consists of all complex linear combinations of products of position variables
$\hat{x}^a$, $a=1,2$, satisfying the commutation relation
\beq
	[\hat{x}^1,\hat{x}^2]= i\theta\ . \label{eq:NC-plane}
\eeq
Here $\theta$ is a constant real number with dimensions of length squared; it controls the ``strength" of the 
noncommutativity of this theory. The algebra $\mathbb{R}^2_{\theta}$ comes equipped with a conjugation operator
``$\dg$" (which one can think of as Hermitian conjugation), and the operators $\hat{x}^a$ are assumed to be invariant
under this operation\footnote{For any complex number $c$ and any $y\in\mathbb{R}^2_{\theta}$ we have
$(c y)^{\dg}= \ov{c} y^{\dg}$, where $\ov{c}$ is the complex conjugate of $c$.}. We see that the algebra 
$\mathbb{R}^2_{\theta}$ is nothing but the universal enveloping 
algebra of the Heisenberg algebra specified by $\hat{x}^a$ and the commutation relation of Eq.~\eqref{eq:NC-plane}. 
The operators $\hat{x}^a$ are sometimes said to be coordinates on a ``noncommutative plane".  In the noncommutative 
theory the notion of a point no longer makes sense, and the smallest area that one can resolve is of order $\theta$. 

In the noncommutative field theory, the notion of integration over space is replaced with a trace in a representation of
the Heisenberg algebra of the noncommutative coordinates $\hat{x}^a$. Usually this representation is taken to be the
Fock representation in which the ladder operators
\beqa
	\hat{a} &=& \frac{1}{\sqrt{2\theta}}(\hat{x}^1 + i\hat{x}^2) \\
	\hat{a}^{\dg} &=& \frac{1}{\sqrt{2\theta}}(\hat{x}^1 - i\hat{x}^2)
\eeqa
act on a Fock space $\mathcal{H}_F$ generated by the action of the raising operator $\hat{a}^{\dg}$ 
on a vacuum state $|0\ran$ which is annihilated by the lowering operator $\hat{a}$. The action functional for the
noncommutative field theory then takes the form
\beq
	S = \int dt\ \text{Tr}_{\mathcal{H}_F}\left\{ (\cdots) \right\}
\eeq
where $(\cdots)$ denotes a Lagrangian written in terms of fields $\hat{\Phi}(t)$ which are
operators on the space $\mathcal{H}_F$, and whose matrix elements are functions of time. 

It is natural to call the formulation of noncommutative field theory that we have just described the ``operator formulation." 
We now describe an alternative formulation, which one might call the ``star-product formulation," which may be more familiar 
to some readers. In this formulation one instead works with fields $\Phi(t,\mb{x})$ which are ordinary functions of the 
coordinates $\mb{x}$ on $\mathbb{R}^2$, but replaces the ordinary product  of functions with the 
\emph{Groenewold-Moyal star product} ``$\star$", which is defined as follows. For any two functions 
$f(\mb{x})$ and $g(\mb{x})$ of $\mb{x}$ we have
\beqa
	f(\mb{x})\star g(\mb{x}) &=& e^{\frac{i}{2} \theta \ep^{ab}\frac{\pd}{\pd y^a}\frac{\pd}{\pd z^a}}f(\mb{y})g(\mb{z})\Big|_{\mb{y}=\mb{z}=\mb{x}} \\
	&=& f(\mb{x})g(\mb{x})+ \frac{i}{2}\theta\ep^{ab}\pd_a f(\mb{x})\pd_b g(\mb{x}) + \dots\ , \nnb
\eeqa  
and where in the last line the ellipses denote terms of order $\theta^2$ and higher. For two functions $f(\mb{x})$ and 
$g(\mb{x})$ which vanish at spatial infinity we have the important property that
\beq
	\int d^2\mb{x}\ f(\mb{x}) \star g(\mb{x})= \int d^2\mb{x}\ f(\mb{x})  g(\mb{x})\ ,
\eeq
which follows after integration by parts on the higher derivative terms in the star product. There is no analogous result for 
integrals of star products of three or more functions.

These two formulations of noncommutative field theory are related by the \emph{Wigner-Weyl} mapping of functions and
operators. This mapping is as follows. Let $f(\mb{x})$ be an ordinary function on $\mathbb{R}^2$ and let
\beq
	\tilde{f}(\mb{k}) = \int d^2\mb{x}\ f(\mb{x}) e^{-i k_a x^a}
\eeq
be its Fourier transform. Then we can define a Weyl-ordered operator $\hat{f}$ by taking the inverse Fourier
transform but replacing $x^a$ with $\hat{x}^a$ in the exponential,
\beq
	\hat{f}= \int\frac{d^2\mb{k}}{(2\pi)^2}\ \tilde{f}(\mb{k}) e^{i k_a \hat{x}^a}\ . \label{eq:Weyl-ordered}
\eeq
One can check that this mapping satisfies the following properties which will be needed later: 
\beqa
	\hat{f}\hat{g} &=& \widehat{f\star g} \\
	\text{Tr}_{\mathcal{H}_F}\left\{\hat{f}\right\} &=& \frac{1}{2\pi\theta}\int d^2 \mb{x}\ f(\mb{x})\ .	
\eeqa
To check the second property one can express the trace over $\mathcal{H}_F$ using a basis $\{|x^1\ran\}$ of eigenstates
of $\hat{x}^1$ as 
\beq
	\text{Tr}_{\mathcal{H}_F}\left\{\hat{f}\right\}= \int_{-\infty}^{\infty}dx^1\ \lan x^1|\hat{f}| x^1\ran
\eeq
and then plug in the expression Eq.~\eqref{eq:Weyl-ordered} for $\hat{f}$.

The Chern-Simons matrix model that we study below is a particular regularization of the NCCS theory in its
operator formulation. Therefore, for our purposes we generally find that the operator formulation of the
NCCS theory is more convenient. However, the star product formulation is still useful for the study of the behavior of the 
theory near the commutative limit $\theta\to 0$, and so we will have occasion to use both formulations of the NCCS theory
in what follows.

\subsection{NCCS theory in the operator formulation}

We now review the operator formulation of the NCCS theory. In the operator formulation, the
NCCS theory consists of three fields
$\hat{X}^a(t)$, $a=1,2$, and $\hat{A}_0(t)$. All fields should be thought of as operators on the Fock space 
$\mathcal{H}_{F}$ whose matrix elements are functions of time. In addition, all fields are Hermitian (i.e., all fields are
invariant under the ``$\dg$" operation on the algebra $\mathbb{R}^2_{\theta}$). 
We also consider the theory on a time interval of length
$T$ and assume periodic boundary conditions in time so that $\hat{X}^a(0)=\hat{X}^a(T),$ and likewise for $\hat{A}_0(t)$. 
In addition to the noncommutativity parameter $\theta,$ the theory includes various coupling constants including $e>0$, an 
electric charge, and $B>0$, a constant magnetic field. We discuss the physical interpretation of this theory as 
representing a charged fluid in a magnetic field later in this section (and we will see that the charge of the particles which 
make up this fluid is actually $q=-e <0$). 

The action for the NCCS theory in the operator formulation takes the form
\beq
	S_{NCCS}= -\frac{eB}{2}\int_0^T dt\ \text{Tr}_{\mathcal{H}_F}\left\{ \ep_{ab} \hat{X}^a D_0\hat{X}^b + 2 \theta \hat{A}_0 \right\}\ , \label{eq:NCCS-action-operator}
\eeq
where we introduced a covariant derivative
\beq
	D_0 \hat{X}^b= \dot{\hat{X}}^b + i[\hat{X}^b,\hat{A}_0]\ .
\eeq
and where the dot denotes a time derivative. The field $\hat{A}_0$ functions as a Lagrange multiplier and its equation of
motion yields the constraint
\beq
	[\hat{X}^1,\hat{X}^2]=i\theta\ . \label{eq:NCCS-constraint}
\eeq
This constraint can only be satisfied by operators $\hat{X}^a$ on an infinite-dimensional space. This
is due to the fact that if the variables $\hat{X}^a$ were finite-dimensional matrices, then the trace of the left-hand side of the
equation is zero while the trace of the right-hand side would be proportional to the size of the matrices.
The CSMM discussed in the next section is a modification of the NCCS theory which features a modified 
constraint that can be satisfied by operators (matrices) on a finite-dimensional space.

If we ignore the term containing $2\theta \hat{A}_0$ for a moment, then 
one can check that the action is invariant under the gauge transformation
\begin{subequations}
\label{eq:NCCS-gauge-trans}
\beqa
	\hat{X}^a &\to& \hat{V}\hat{X}^a \hat{V}^{\dg} \\
	\hat{A}_0 &\to& \hat{V}\hat{A}_0 \hat{V}^{\dg} + i \hat{V} \dot{\hat{V}}^{\dg},
\eeqa
\end{subequations}
where $\hat{V}(t)$ is an arbitrary time-dependent unitary operator on the Fock space $\mathcal{H}_F$. In particular,
this follows from the fact that, under this transformation, the covariant derivative transforms as 
$D_0 \hat{X}^b \to \hat{V} D_0 \hat{X}^b \hat{V}^{\dg}$. To understand these gauge transformations in the presence of 
the term $2\theta \hat{A}_0$, we need to constrain the allowed $\hat{V}$'s that we consider~\cite{PN}. To motivate 
this restriction we now briefly discuss some aspects of the geometry of the noncommutative plane.

Consider the occupation number basis $\{|n\ran\}_{n\in\mathbb{N}}$ of the Fock space $\mathcal{H}_F$ 
($|n\ran \propto (\hat{a}^{\dg})^n|0\ran$). The
radius squared operator $\hat{R}^2= \delta_{ab}\hat{x}^a\hat{x}^b$ is diagonal in this basis and we have 
$\hat{R}^2|n\ran= 2\theta(n+\frac{1}{2})|n\ran$. Thus, the occupation number $n$ can be
identified with the distance squared from the origin in the noncommutative plane. We now restrict our attention to gauge 
transformations defined by unitary operators $\hat{V}(t)$ which act as the identity 
on states $|n\ran$ with $n$ sufficiently large, say $n>N_0$. The actual value of $N_0$ is not important for the argument.
This is the noncommutative analogue of requiring gauge transformations in a commutative gauge theory on
the space $\mathbb{R}^2$ to tend to the identity at spatial infinity. 

With this restriction on possible gauge transformations, the unitary operator $\hat{V}(t)$ defines a map from the 
periodic time interval $[0,T)$ to $U(N_0)$, the group
of unitary matrices of size $N_0$. Large gauge transformations are those $\hat{V}(t)$ which correspond to a nontrivial element
of the homotopy group $\pi_1(U(N_0))=\mathbb{Z}$. The full NCCS action is not invariant under these large gauge
transformations because of the presence of the $2\theta \hat{A}_0$ term. 
In Ref.~\onlinecite{PN}, Polychronakos and Nair have 
shown that requiring the exponential $e^{i\frac{S_{CSMM}}{\hbar}}$ to be invariant under these large
gauge transformations enforces a 
quantization rule on $\theta$ which states that
\beq
	e B\theta= \hbar m,\ m\in\mathbb{Z}\ ,
\eeq
or
\beq
	\theta= \ell_B^2 m\ ,\ m\in\mathbb{Z}\ , \label{eq:level-quantization}
\eeq
where $\ell_B^2= \frac{\hbar}{eB}$ is the square of the magnetic length defined earlier. This quantization rule is the
noncommutative analogue of the level quantization which obtains in ordinary (say $SU(N)$) Chern-Simons theory on a 
commutative space.

\subsection{NCCS theory in the star product formulation}

We now discuss the NCCS theory in the star product formulation. In this form the theory looks very similar to the
ordinary Chern-Simons theory (i.e., Chern-Simons theory on the commutative space $\mathbb{R}^2$). We proceed by deriving the star
product formulation of the NCCS theory from the operator formulation by using the Wigner-Weyl mapping discussed earlier
in this section. To do this we need to know how spatial derivatives are represented in the operator formulation of the
theory. Derivative operators $\hat{\pd}_a$ in the operator formulation of noncommutative field theory are defined by
\beq
	\hat{\pd}_1= \frac{i\hat{x}^2}{\theta}\ ,\ \hat{\pd}_2=-\frac{i\hat{x}^1}{\theta}\ 
\eeq
and one can check that
\beq
	[\hat{\pd}_a,\hat{x}^b]=\delta_a^b\ ,
\eeq
just as one has for ordinary derivatives of functions on $\mathbb{R}^2$. In addition, in the Wigner-Weyl mapping one has
\beq
	[\hat{\pd}_a,\hat{f}]= \widehat{\pd_a f}\ ,
\eeq
so under this mapping the ordinary derivative of a function $f(\mb{x})$ with respect to $x^a$ is mapped to the commutator
of $\hat{\pd}_a$ with $\hat{f}$ (i.e., the \emph{adjoint} action of $\hat{\pd}_a$ on $\hat{f}$).

The first step towards deriving the star product formulation of NCCS theory is to make a change of variables in the operator
formulation by defining two new fields $\hat{A}_a$, $a=1,2$, which are related to the fields $\hat{X}^a$ by
\beq
	\hat{X}^a= \hat{x}^a + \theta \ep^{ab} \hat{A}_b\ .
\eeq
Under a gauge transformation the new fields transform as\footnote{This is derived by requiring the gauge transformation
of $\hat{x}^a+\theta \ep^{ab}\hat{A}_b$ to coincide with the gauge transformation of $\hat{X}^a$ from 
Eq.~\eqref{eq:NCCS-gauge-trans}.}
\beq
	\hat{A}_a\to \hat{V}\hat{A}_a\hat{V}^{\dg} + i\hat{V}[\hat{\pd}_a,\hat{V}^{\dg}]\ .
\eeq
This transformation resembles the transformation of an ordinary non-Abelian gauge field. In addition, in the
new variables, the NCCS constraint of Eq.~\eqref{eq:NCCS-constraint} becomes
\beq
	\hat{F}_{12}= 0\ ,
\eeq
where we defined the field strength for noncommutative gauge theory as
\beq
	\hat{F}_{ab}= [\hat{\pd}_a,\hat{A}_b]-[\hat{\pd}_b,\hat{A}_a]- i[\hat{A}_a,\hat{A}_b]\ . 
\eeq
Thus, the constraint in NCCS theory is an exact noncommutative analogue of the constraint enforced by
the temporal component of the gauge field in ordinary Chern-Simons theory on a commutative space.

After tedious algebra (including many uses of the cyclic property of the trace) 
one can show that after performing this transformation the NCCS action takes the form
\begin{align}
	S_{NCCS} = -\frac{eB\theta^2}{2}&\int_0^T dt\ \text{Tr}_{\mathcal{H}_F}\Bigg\{ \ep^{ab}\hat{A}_a\dot{\hat{A}}_b - \ep^{ab} \hat{A}_0 [\hat{\pd}_a,\hat{A}_b] \nnb \\
 +\ \ep^{ab} \hat{A}_b&[\hat{\pd}_a,\hat{A}_0] + \frac{2i}{3}\ep^{\mu\nu\lam}\hat{A}_{\mu}\hat{A}_{\nu}\hat{A}_{\lam}\Bigg\}\ ,
\end{align}
where the Greek indices $\mu,\nu,\lam$ run over the range $0,1,2$. There is one subtle point in the derivation of
this equation which involves a term which is a total time derivative. Specifically, after the transformation from the 
$\hat{X}^a$ variables to the $\hat{A}_a$ variables one finds a term
\beq
	-\frac{eB}{2}\int_0^T dt\ \text{Tr}_{\mathcal{H}_F}\left\{ -\theta \hat{x}^a \dot{\hat{A}}_a\right\}\ .
\eeq
Since $\hat{x}^a$ has no time dependence this term is a total derivative, and it evaluates to zero since we assumed periodic
boundary conditions on all fields in the time direction. 

Finally, we apply the Wigner-Weyl mapping to write the NCCS action in the star product formulation as
\begin{align}
	S_{NCCS}= \frac{eB\theta}{4\pi}\int_0^T dt \int d^2\mb{x}\ &\ep^{\mu\nu\lam} \Big( A_{\mu}\star \pd_{\nu}A_{\lam} \nnb \\
-&\frac{2i}{3}A_{\mu}\star A_{\nu}\star A_{\lam}  \Big)\ .
\end{align}
The quantization condition on $\theta$ (Eq.~\eqref{eq:level-quantization}) then implies that the coefficient of the action is 
\beq
	\frac{eB\theta}{4\pi}= \frac{\hbar m}{4\pi}\  .
\eeq
Then, in units where $\hbar=1,$ we find the NCCS action at level $m\in\mathbb{Z}$. If we take $\ell_B^2\to 0$, 
which also sends $\theta \to 0$, then we recover the ordinary $U(1)$ Chern-Simons theory at level $m$ (again
with $\hbar=1$ for now),
\beq
	S_{CS}= \frac{m}{4\pi}\int_0^T dt \int d^2\mb{x}\ \ep^{\mu\nu\lam}A_{\mu}\pd_{\nu}A_{\lam}\ .
\eeq
For completeness we note here that in the star product formulation the noncommutative field strength is
\beq
	F_{\mu\nu}= \pd_{\mu}A_{\nu} - \pd_{\nu}A_{\mu} - i(A_{\mu}\star A_{\nu} - A_{\nu}\star A_{\mu})\ ,
\eeq
and the equation of motion of the NCCS theory is equivalent to $F_{\mu\nu}=0$, just like in ordinary Chern-Simons
theory. 

\subsection{Fluid interpretation of the NCCS theory at small $\theta$}

We now discuss the behavior of the NCCS theory in the limit of weak noncommutativity in which $\theta$ is assumed to 
be small. Note that since $\theta$ has units, and since there is no other length scale in the problem to compare $\theta$
to, it is more accurate to say that in this section we study a \emph{truncation} of the NCCS theory at first order
in $\theta$. In the star product formulation of the theory this truncation simply amounts to neglecting terms of order
$\theta^2$ and higher in the star product of functions.
In this limit we will see that the NCCS theory has an 
interpretation as describing a fluid of charged particles in 
a constant magnetic field $B$, as was discussed by Susskind~\cite{susskind} (see also Refs.~\onlinecite{jackiw2002,P2}). 

To consider the NCCS theory in the regime of small $\theta$ we start by using the cyclic property of the trace
to write the action in the form 
\begin{align}
	S_{NCCS}= -\frac{eB}{2}\int_0^T dt&\ \text{Tr}_{\mathcal{H}_F}\Bigg\{ \ep_{ab}\hat{X}^a\dot{\hat{X}}^b \nnb \\
+&\ 2\hat{A}_0 \left(\theta + i[\hat{X}^1,\hat{X}^2]\right)\Bigg\}\ .
\end{align}
We then use the Wigner-Weyl mapping, and keep only the terms up to order $\theta$ in the star product, to find that in the
limit of small $\theta$
\begin{align}
	S_{NCCS} \to -\frac{eB}{2}\frac{1}{2\pi\theta}&\int_0^T dt \int d^2\mb{x}\ \Big( \ep_{ab}X^a\dot{X}^b \nnb \\
+ &2\theta A_0 \left(1-\ep^{ab}\pd_a X^1 \pd_b X^2\right)\Big)\ . \label{eq:NCCS-small-theta}
\end{align}
%where for any $f(\mb{x})$ and $g(\mb{x})$ we have the Poisson bracket
%\beq
%	\{f,g\}= \ep^{ab}\pd_a f\pd_b g\ .
%\eeq

Susskind observed that in this limit the NCCS theory describes the dynamics of a charged fluid at constant density
$\rho_0=\frac{1}{2\pi \theta}$ in a 
constant magnetic field $B$, and in the limit where the cyclotron frequency is sent to infinity. In fact, in Susskind's original 
derivation he starts with the fluid description and then observes that it coincides with the small $\theta$ limit of the 
NCCS theory. 
%Therefore our discussion of the NCCS theory is in the reverse order to the discussion in Ref.~\onlinecite{susskind}. 
We now briefly remind the reader of this connection between the NCCS theory and fluid dynamics.

The starting point is the \emph{Lagrange} description\footnote{The relation between noncommutative gauge theory
and the Lagrange description of a fluid is discussed in detail in Ref.~\onlinecite{jackiw2002}.} 
of a fluid of charged particles moving on the plane $\mathbb{R}^2$ in a 
background electromagnetic field. In the Lagrange
description of a fluid one keeps track of the motion of the individual particles in the fluid, and measures their current position
with respect to some reference configuration. In this description we use coordinates $\mb{x}$ to describe the reference
configuration of the fluid and coordinates $X^a(t,\mb{x})$, $a=1,2$, to describe the configuration of the fluid at a later time
$t$. Without loss of generality, we may assume that $X^a(0,\mb{x})=x^a$. Thus, $X^a(t,\mb{x})$ is the position, at time
$t$, of the fluid particle which was at position $x^a$ at $t=0$. We also assign a constant density $\rho_0$ to the fluid in 
the reference configuration.

The action for a Lagrange fluid made up of particles of mass $M$ and charge $q$ in the presence of a background 
electromagnetic field takes the form
\begin{align}
	S = \int_0^T dt\int d^2\mb{x}\ \rho_0 \Bigg(\frac{1}{2}M \delta_{ab} \dot{X}^a&\dot{X}^b + q \mathcal{A}_a(t,\mb{X})\dot{X}^a  \nnb \\ 
  -q &\vphi(t,\mb{X})  \Bigg)\ , \label{eq:charged-fluid}
\end{align}
where $\mathcal{A}_a(t,\mb{X})$ and $\vphi(t,\mb{X})$ are the vector and scalar potentials, respectively,
for the external electromagnetic
field. Intuitively, this action is just the sum over all particles in the fluid of the 
ordinary action for a massive charged particle in a background electromagnetic field. However, the discrete sum over particle 
labels has been replaced with an integration over the reference coordinates $\mb{x}$ weighted with the density $\rho_0$ in 
the reference configuration. The reference coordinates $\mb{x}$ can therefore be considered as a set of continuous particle
labels.

To see the connection of the fluid model to the NCCS theory we first
place the system in a uniform background magnetic field with strength $B>0$. 
This can be accomplished by setting $\vphi(t,\mb{X})=0$ and
\beq
	\mathcal{A}_a(t,\mb{X})= -\frac{B}{2}\ep_{ab}X^b\ ,
\eeq
where we have chosen the symmetric gauge for the vector potential. Next, we set the mass of the particles to zero, 
$M=0$. This corresponds to taking the cyclotron frequency $\omega_c=\frac{e B}{M}$ to infinity, which is similar to a 
projection into the lowest Landau level (since $\hbar\omega_c$ is the energy gap between Landau levels). Finally, we
take the charge of the particles to be $q=-e$ with $e>0$. Then at this point the action reads as
\beq
	S= -\frac{e B}{2} \rho_0 \int_0^T dt\int d^2\mb{x}\ \ep_{ab}X^a\dot{X}^b\ .
\eeq
Note that $\rho_0$ can be pulled out of the integral since we assumed it was constant.
We also mention here that
our conventions for the direction of the magnetic field and the charge of the particles in the fluid exactly matches our 
conventions for the setup of the quantum Hall problem from Sec.~\ref{sec:GC-hall-visc}.
%We also mention here that 
%our convention in which the particles carry charge $e>0$ and the magnetic field $B>0$ points in the positive 
%$z$ direction is equivalent to choosing the particles to have charge $-e<0$ (i.e., electrons) and placing those
%particles in a magnetic field of magnitude $B$ but pointing in the minus $z$ direction. 

The next step is
to incorporate a Lagrange multiplier which enforces the constraint that the fluid remains at the constant density
$\rho_0$ at all times. The density $\rho(t,\mb{X})$ of the fluid at time $t$ is related to the initial density $\rho_0$ by the 
Jacobian $\ep^{ab}\pd_a X^1 \pd_b X^2$ of the map from the reference coordinates
to the fluid coordinates $\mb{X}$ at time $t$ as
\beq
	\rho(t,\mb{X}) \ep^{ab}\pd_a X^1 \pd_b X^2= \rho_0\ ,
\eeq
where we remind the reader that $\pd_a$ is a shorthand for $\frac{\pd}{\pd x^a}$, i.e., a derivative with respect to the
reference coordinates $x^a$. Then the constraint that $\rho(t,\mb{X})=\rho_0$ for all $t$ can be written as
\beq
	\ep^{ab}\pd_a X^1 \pd_b X^2=1.
\eeq
% in terms of a Poisson bracket as
%\beq
%	\{X^1,X^2\}= 1\ .
%\eeq
We denote the Lagrange multiplier enforcing this constraint by $A_0(t,\mb{x})$, and write the action with the constraint
included in the form 
\begin{align}
	S= -\frac{e B}{2} \rho_0 \int_0^T dt&\int d^2\mb{x}\ \Bigg(\ep_{ab}X^a\dot{X}^b  \nnb \\
+&\ 2\theta A_0 \left( 1 - \ep^{ab}\pd_a X^1 \pd_b X^2\right)\Bigg)\ ,
\end{align}
where we have introduced a parameter $\theta$ with units of (length)$^2$. With this choice, the Lagrange multiplier field 
$A_0$ has units of (time)$^{-1}$. 

We can now see that the small $\theta$ limit of the NCCS action from Eq.~\eqref{eq:NCCS-small-theta} is exactly the
action for a fluid of particles with charge $q=-e$ at the constant density $\rho_0=\frac{1}{2\pi\theta}$ in a constant 
background magnetic
field $B$ in the limit in which the cyclotron frequency is taken to infinity. This limit is analogous to the projection into the
lowest Landau level, and it is the physical reason why this fluid theory (and the NCCS theory) is expected to describe
FQH physics in the lowest Landau level~\cite{susskind}. In the full NCCS theory we should then interpret the fields
$\hat{X}^a(t)$ as describing the positions of particles in a fluid on a noncommutative space, as discussed by 
Susskind~\cite{susskind} (see also Ref.~\onlinecite{P2} for a review of the physics of 
such \emph{noncommutative fluids}).

\section{The Chern-Simons Matrix Model}
\label{sec:CSMM}

In this section we discuss the Chern-Simons matrix model (CSMM), which was introduced
by Polychronakos in Ref.~\onlinecite{P1}. This model can be thought of as a particular regularization of the 
operator formulation of the NCCS theory, in which the fields $\hat{X}^a(t)$ (which were
operators on the \emph{infinite-dimensional} Fock space $\mathcal{H}_F$) are now finite $N\times N$ matrices $X^a(t)$
instead. Note that we \emph{do not} use a hatted notation for the finite size matrix variables of the NCCS theory. 
The parameter $N$ serves as a regulator which should be taken to infinity to recover the NCCS theory discussed
in the previous section. The fluid interpretation of the NCCS theory carries over to the CSMM, so we still interpret the matrix
variables $X^a(t)$ as representing the coordinates of particles in a fluid on a noncommutative space, only now the fluid turns
out to occupy a finite area of this space. In other words, the CSMM is a model of a finite droplet \emph{droplet} of 
noncommutative fluid.

Since the CSMM can be difficult to understand, we begin this section by making a few
remarks about our notation and conventions, and then discuss some subtleties of this model. 
We then review the quantization of this model following Refs.~\onlinecite{P1,HVR}. Finally, we review (following
the original discussion in Ref.~\onlinecite{P1}) the calculation 
of the area $A$ and density $\rho_0$ of the droplet of noncommutative fluid represented by the ground state of the CSMM.
We will then be able to identify the CSMM having $\theta=\ell_B^2 m$ as describing the $\nu=\frac{1}{m}$ Laughlin state
by comparing the results for $\rho_0$ and $A$ to the known answers for a droplet of FQH fluid in the $\nu=\frac{1}{m}$
Laughlin state in the limit of a large number of particles $N$. 

\subsection{Some remarks on notation}

The CSMM, and especially the quantization of this model, can be quite tricky due to two 
separate noncommutative structures which appear. First, at the classical level the degrees of freedom in this model are 
Hermitian $N\times N$ matrix variables $X^1$, $X^2$,
$A_0$, as well as a complex vector $\Psi$ of length $N$. All of these variables are functions of time. Since some of the 
variables are matrix variables, ordinary (i.e., classical) matrix multiplication of these variables is 
not commutative. Next, upon quantization of the model, the \emph{matrix elements} of $X^1$, $X^2$, and $A_0$ 
(and also the components of $\Psi$) become
\emph{operators} on a separate Hilbert space, which is unrelated to the vector space on which the classical matrix variables 
act. Thus, in the quantized matrix model there are two sources of noncommutativity. The
first source is the fact that we are dealing with matrix variables from the start, and the second source comes from the
fact that the matrix elements of the original matrix variables are now operators on a second Hilbert space, and so multiplication
of individual matrix elements does not commute either, but for a different reason. 

In an attempt to present this model in as clear a manner as possible, we will adhere to the following notational 
conventions. First, we use $[\cdot,\cdot]_M$ to denote a matrix commutator of classical matrices, and use 
$[\cdot,\cdot]$ (with no subscript) to denote the commutator of quantum operators. 
We also reserve the symbol $\dg$ to denote Hermitian conjugation of quantum operators. In all manipulations
with classical matrix variables, we instead use an overline to denote complex conjugation of a matrix and a superscript `T' to 
denote a transpose. So if $A$ is an $N\times N$ matrix variable, then $\ov{A}^T$ is its transpose conjugate, i.e., if 
$A$ has matrix elements $A_{jk}$, then the matrix elements of $\ov{A}^T$ are $(\ov{A}^T)_{jk}=\ov{A}_{kj}$ (and
Hermitian matrices satisfy the relation $\ov{A}^T=A$). As we 
mentioned before, in the quantum theory the matrix elements $A_{jk}$ are promoted to operators on a Hilbert
space. We denote the Hermitian conjugate (with respect to the inner product on this Hilbert space) of the operator $A_{jk}$
by $A^{\dg}_{jk}$. Note that for a generic matrix variable $A$ it is entirely possible that the operator $A^{\dg}_{jk}$
is not the same as the operator $(\ov{A}^T)_{jk}$. In what follows we also make every effort to avoid using `$i$' as an index,
and instead try to reserve it for the symbol meaning $\sqrt{-1}$, and occasionally for the differential geometry operation 
$i_{\un{v}}$ of interior multiplication by a vector field $\un{v}$.

\subsection{Description of the model}

In this subsection we describe the CSMM of the Laughlin quantum Hall states~\cite{P1}. 
The degrees of 
freedom in this model are two $N\times N$ matrices $X^a(t)$, $a=1,2$, an $N\times N$ matrix $A_0(t)$, and a complex
vector $\Psi(t)$ of length $N$. All degrees of freedom depend on time. The matrices $X^a(t)$ and $A_0(t)$ are all Hermitian
and so they have real eigenvalues. The variables $X^a$ are to be interpreted as 
coordinates in the Lagrange description of a fluid on the noncommutative plane, in accordance with the physical ideas of 
Susskind and Polychronakos~\cite{susskind,P1}(and as we reviewed at the end of Sec.~\ref{sec:NCCS}). 
The number $N$ will later be identified with
the number of electrons in a Landau level. The action for the CSMM takes the form
\begin{align}
	S_{CSMM} &= -\frac{eB}{2}\int_0^T dt\ \text{Tr}\Big\{ \ep_{ab}X^a D_0 X^b + 2\theta A_0 \nnb \\
+&\ \tomega \delta_{ab}X^a X^b \Big\} + \int_0^T \ov{\Psi}^{T}(i\dot{\Psi}+ A_0 \Psi)\ ,
\end{align}
where
\beqa
	D_0 X^b &:=& \dot{X}^b + i[X^b,A_0]_M \nnb \\
	&=& \dot{X}^b- i[A_0,X^b]_M
\eeqa
is a covariant derivative. Here we view $\Psi$ as a column vector and $\ov{\Psi}^T$ denotes the row vector whose
elements are the complex conjugates of the elements of $\Psi$. 
In addition, $e$ and $B$ are the same charge and constant magnetic field from Sec.~\ref{sec:NCCS}, $\tomega$ is a 
frequency (the term with $\tomega$ is a quadratic potential for the noncommutative coordinates $X^a$), and $\theta$ is a 
parameter with units of length squared. We assume periodic boundary conditions on all the fields in the time direction,
for example $X^a(0)=X^a(T)$, so that the time direction is a circle of circumference $T$. Note that the action as written here 
differs slightly in the details (signs, etc.) from Ref.~\onlinecite{P1}, but is consistent with our interpretation of this model
and the NCCS theory as describing a noncommutative fluid of particles with charge $-e<0$.

At this point we would like to emphasize that the frequency $\td{\omega}$ appearing in the parabolic potential
term of the CSMM has no relation to the
cyclotron frequency $\omega_c=\frac{e B}{M}$ in the quantum Hall problem. Indeed, as we discussed in
Sec.~\ref{sec:NCCS}, the NCCS theory (and therefore the CSMM as well) describes a charged fluid in a magnetic field
in the limit in which the mass $M$ of the particles making up the fluid has been sent to zero. This sends the cyclotron
frequency $\omega_c$ to infinity. Therefore, the CSMM contains no information related to the cyclotron frequency or
the energy of a Landau level.

We now discuss the gauge symmetry in the CSMM.
If we ignore the term with $2\theta A_0$ for a moment, then we can see that the rest of the action is invariant under a
$U(N)$ gauge transformation
\begin{subequations}
\beqa
	X^a &\to& VX^a \ov{V}^T \\
	A_0 &\to& VA_0 \ov{V}^T + i V \dot{\ov{V}}^T \\
	\Psi &\to& V\Psi\ ,
\eeqa
\end{subequations}
where $V(t)$ is an arbitrary time-dependent $U(N)$ matrix. The presence of the term $2\theta A_0$ in the action means that
the action is not invariant under large $U(N)$ gauge transformations which are maps from $[0,T) \to U(N)$
which correspond to a nontrivial element in the homotopy group $\pi_1(U(N))=\mathbb{Z}$. Since we would like 
$e^{i\frac{S_{CSMM}}{\hbar}}$ to be invariant under any gauge transformation, these large gauge transformations enforce a 
quantization rule on $\theta$ (the argument is identical to the argument for the full NCCS theory from Sec.~\ref{sec:NCCS}) 
which states that
\beq
	e B\theta= \hbar m,\ m\in\mathbb{Z}\ ,
\eeq
or
\beq
	\theta= \ell_B^2 m\ ,\ m\in\mathbb{Z}\ . 
\eeq

The gauge field $A_0$ can be interpreted as a matrix Lagrange multiplier. If we look at the equation of motion resulting from a 
variation of $A_0$, then we find that $A_0$ enforces the constraint
\beq
	ieB[X^1,X^2]_M+eB\theta\mathbb{I} - \Psi\ov{\Psi}^T= 0\ . \label{eq:constraint}
\eeq
This constraint should be compared with Eq.~\eqref{eq:NCCS-constraint} for the NCCS theory. In the NCCS case the
contribution from the vector $\Psi$ is absent and the constraint can only be realized by infinite-dimensional matrices (i.e.,
operators on $\mathcal{H}_F$).
It is the presence of the vector $\Psi$ which allows this constraint to be 
realized by finite-dimensional matrices, and this is why the CSMM can be thought of as a regularization of the NCCS
theory. We refer the reader to Ref.~\onlinecite{P1} for the detailed analysis of the constraint in the classical solution of 
the CSMM (which is also closely related to the Calogero model of interacting particles in one spatial dimension).
In this paper our main focus is on the solution of the model in the quantum case. 

We now make a few remarks and set up some notation relating to the transformation properties of the fields under the
action of the group $U(N)$. The field $\Psi$ transforms in the fundamental representation of $U(N)$. We indicate this
by writing the components of $\Psi$ with an upper Latin index, $\Psi^j$, $j=1,\dots,N$. Under a $U(N)$ transformation we 
have
\beq
	\Psi^j \to {V^j}_k \Psi^k\ ,
\eeq
where ${V^j}_k $ are the matrix elements of a unitary matrix $V$ in $U(N)$. Next, the transpose conjugate $\ov{\Psi}^T$
transforms in the anti-fundamental representation of $U(N)$, $\ov{\Psi}^T\to \ov{\Psi}^T\ov{V}^T$. We indicate this
by writing the components of $\ov{\Psi}^T$ with a lower index, $\ov{\Psi}_j$, $j=1,\dots,N$ (and recall that the components 
of $\ov{\Psi}^T$ are just the complex conjugates of the components of $\Psi$). In components we have
\beq
	\ov{\Psi}_j \to \ov{\Psi}_k {(\ov{V}^T)^k}_j\ .
\eeq
Finally, the matrix variables $X^a$ transform in the adjoint representation of $U(N)$, $X^a\to V X^a \ov{V}^T$. Thus, the
index structure of $X^a$ is such that it has one upper and one lower index, ${(X^a)^j}_k$, $j,k=1,\dots,N$. The component
form of the $U(N)$ transformation is then
\beq
	{(X^a)^j}_k \to {V^j}_{\ell} {(X^a)^{\ell}}_{m} {(\ov{V}^T)^m}_k\ .
\eeq
These conventions will be extremely useful later when we try to write down quantum states that respect the constraint
of the CSMM. 

We already mentioned that the matrix variables $X^a$ are Hermitian matrices. Thus, their matrix elements ${(X^a)^j}_k$
are generically complex numbers. For the quantization of this system it will be more convenient to parametrize $X^a$
in terms of scalar variables which are manifestly real. Then, when we quantize the theory, these real variables will be
promoted to Hermitian operators on the quantum Hilbert space. Our choice of parametrization is as follows. First, let 
$T^A$, $A=1,\dots,N^2-1$, be the $N\times N$ generators, in the fundamental representation, of the Lie algebra of $SU(N)$. 
The matrices $T^A$ are all Hermitian and traceless, and can be normalized to obey the relations
\begin{subequations}
\beqa
	\text{Tr}\{T^A T^B\}&=& \delta^{AB} \\
	\left[T^A,T^B\right]_{M} &=& i \sum_{C=1}^{N^2-1} f^{ABC} T^C\ ,
\eeqa
\end{subequations}
where $f^{ABC}$ are the structure constants for $SU(N)$. These structure constants have a very important property which
is that they are antisymmetric under exchange of \emph{any} two indices $A,B$, or $C$ (typically one only expects
antisymmetry under $A \leftrightarrow B$). We will take advantage of this property later on. Using the generators
$T^A,$ we can parametrize $X^a$ (for $a=1,2$) as
\beq
	X^a(t)= x^a_0(t)\frac{\mathbb{I}}{\sqrt{N}} + \sum_{A=1}^{N^2-1} x^a_A(t) T^A\ , \label{eq:expansion}
\eeq
where $x^a_0(t)$ and $x^a_A(t)$, $A=1,\dots,N^2-1$, are $N^2$ real scalar variables. In the quantum theory these 
variables will be promoted to Hermitian operators. The factor of $\sqrt{N}$ on the identity matrix term was chosen for 
convenience. 

The Poisson brackets for this system can be obtained from the corresponding symplectic form, which can in turn be read off
from the action (which is first order in time derivatives). The full symplectic form on the phase space for this system
is
\beq
	\Omega= \Omega_{X}+\Omega_{\Psi}
\eeq
with 
\beq
	\Omega_{X}= -e B\sum_{A=0}^{N^2-1} dx^1_A\wedge dx^2_A
\eeq
%\beq
%	\Omega_{X}= eB\text{Tr}\{dX^1\wedge dX^2\} = e B {(dX^1)^j}_k \wedge {(dX^2)^k}_j 
%\eeq
and
\beq
	\Omega_{\Psi}= -i\ d\Psi^j\wedge d\ov{\Psi}_j\ .
\eeq
Our conventions for Poisson brackets are as follows. To any function $f$ on phase space we associate a vector field
$\un{v}_f$ defined as the solution to the equation $df= -i_{\un{v}_f}\Omega$. Then the Poisson bracket of any two functions
$f$ and $g$ is given by $\{f,g\}= i_{\un{v}_f}i_{\un{v}_g}\Omega$. Using this convention we obtain the classical
Poisson brackets (with $A,B=0,\dots,N^2-1$ now)\footnote{The reader should beware that the symbol $B$ is now being used
for two purposes. It is the strength of the magnetic field felt by the noncommutative fluid described by the CSMM and NCCS 
theory, and it is also (along with the capital Latin letters $A,C,\dots$) an index on the $SU(N)$ generators 
$T^A$ and the variables $x_A$. It should be clear from the context whether $B$ represents the magnetic field strength or an
index.}
\begin{subequations}
\beqa
	\{ x^1_A ,x^2_B\} &=& \frac{1}{eB}\delta_{AB} \label{eq:X-poisson-brackets}\\
%\{ {(X^1)^j}_k, {(X^2)^{\ell}}_m\} &=& \frac{1}{eB}\delta^j_m \delta^{\ell}_k \\
	\{ \Psi^j, \ov{\Psi}_k\} &=& -i\delta^j_k\ .
\eeqa
\end{subequations}
Upon quantization, in which we replace Poisson brackets with commutators as $\{ f,g\} \to -\frac{i}{\hbar}[f,g]$, we find
the commutation relations in the quantum CSMM to be
\begin{subequations}
\label{eq:CSMM-comm-rels}
\beqa
	[x^1_A ,x^2_B] &=& i\ell_B^2\delta_{AB}  \\
%\left[ {(X^1)^j}_k, {(X^2)^{\ell}}_m \right] &=& i\ell_B^2\delta^j_m \delta^{\ell}_k \\
	\left[ \Psi^j, \ov{\Psi}_k \right] &=& \hbar \delta^j_k\ ,
\eeqa
\end{subequations}
where $\ell_B^2=\frac{\hbar}{eB}$ is the magnetic length.

Finally, when the gauge field $A_0$ is set to zero, the Hamiltonian for this system is given by
\beq
	H_{CSMM}= \frac{eB\tomega}{2}\text{Tr}\{\delta_{ab}X^a X^b\}\ .
\eeq
All of the energy in the system is associated with the harmonic trap, and the only energy scale is associated with frequency
$\tomega$ of the harmonic trap. 

We now review the quantization of this model.

\subsection{Quantization of the CSMM}

We now discuss the quantization of the CSMM. Instead of trying to solve the constraint before quantization, we follow
previous approaches to this model and first quantize, then impose the constraint on quantum states, i.e., physical states should
be annihilated by the constraint operator. As we discussed above, upon quantization the matrix elements of $X^1$
and $X^2$ and the components of $\Psi$ obey the quantum commutation relations from Eq.~\eqref{eq:CSMM-comm-rels}.
%\beqa
%	[x^1_A ,x^2_B] &=& i\ell_B^2\delta_{AB}  \\
%%\left[ {(X^1)^j}_k, {(X^2)^{\ell}}_m \right] &=& i\ell_B^2\delta^j_m \delta^{\ell}_k \\
%	\left[ \Psi^j, \ov{\Psi}_k \right] &=& \hbar \delta^j_k\ .
%\eeqa
In what follows we instead work with the oscillator variables
\beq
	b^j= \frac{1}{\sqrt{\hbar}}\Psi^j \ ,
\eeq
with $b^{\dg}_j= \frac{1}{\sqrt{\hbar}}\ov{\Psi}_j$, and
\beq
	a_A = \frac{1}{\ell_B\sqrt{2}}( x^1_A + i x^2_A)\ ,
\eeq
with $a^{\dg}_A = \frac{1}{\ell_B\sqrt{2}}( x^1_A - i x^2_A)$. These variables obey the commutation relations
\beqa
	\left[a_A, a^{\dg}_B\right] &=& \delta_{AB} \\
	\left[ b^j, b^{\dg}_k\right]&=& \delta^j_k\ .
\eeqa

The Hamiltonian for this system has the form
\beqa
	H_{CSMM} &=&\frac{eB\tomega}{2}\delta_{ab} {(X^a)^j}_k {(X^b)^k}_j \nnb \\
	&=& \frac{eB\tomega}{2}\sum_{A=0}^{N^2-1}\delta_{ab} x^a_A x^b_A\ .
\eeqa
In terms of the oscillator variables $a_A$ and $a^{\dg}_A$ this becomes
\beq
	H_{CSMM}= \hbar\tomega\frac{N^2}{2} + \hbar\tomega \sum_{A=0}^{N^2-1}a^{\dg}_A a_A\ .
\eeq
Note that the first term represents the zero point energy of $N^2$ harmonic oscillators.

Next we turn to an analysis of the constraint. Classically, and in terms of the variables $x^a_A$, the constraint
from Eq.~\eqref{eq:constraint} takes the form
\beq
	-eB \sum_{A,B,C=1}^{N^2-1}x^1_A x^2_B f^{ABC} T^C + eB \theta \mathbb{I} - \Psi \ov{\Psi}^T= 0\ .
\eeq
To interpret the constraint in the quantum theory we study its $j,k$ matrix element
\beq
	-eB \sum_{A,B,C=1}^{N^2-1}x^1_A x^2_B f^{ABC} {(T^C)^j}_k + eB \theta \delta^j_k - \Psi^j \ov{\Psi}_k = 0\ .
\eeq
In terms of the oscillator variables one can show that this matrix element of the constraint takes the form 
\begin{align}
	i\frac{\hbar}{2}\sum_{A,B,C=1}^{N^2-1}(a^{\dg}_A a_B + a_B a^{\dg}_A)f^{ABC}&{(T^C)^j}_k  \nnb \\
+&\  eB \theta \delta^j_k - \hbar b^j b^{\dg}_k = 0\ .
\end{align}
Note that in deriving this expression we needed to use the antisymmetry of the structure constants $f^{ABC}$ under 
exchange of its indices. Finally, we use the commutation relations of the oscillator variables to rewrite this as
\beq
	i\hbar\sum_{A,B,C=1}^{N^2-1}a^{\dg}_A a_B f^{ABC}{(T^C)^j}_k +  (eB \theta-\hbar) \delta^j_k - \hbar b^{\dg}_k b^j = 0\ ,
\eeq
where we used the fact that $\sum_{A,B=1}^{N^2-1} \delta_{AB} f^{ABC}=0$. Note the shift in the coefficient of the
$\delta^j_k$ term which resulted from this manipulation\footnote{In Ref.~\onlinecite{P1} Polychronakos instead
performs normal-ordering of the constraint by making the \emph{replacement} $b^j b^{\dg}_k \to b^{\dg}_k b^j$. There is 
then no shift of the coefficient of the $\delta^j_k$ term. This difference between normal-ordering the constraint vs. treating
it \emph{as is} completely accounts for the fact that Polychronakos found that the CSMM with $\theta=\ell_B^2 m$ describes 
the $\nu=\frac{1}{m+1}$ Laughlin state, while we will find that it describes the $\nu=\frac{1}{m}$ Laughlin state (if
we treated the constraint like Polychronakos then this would result in a trivial replacement of $m\to m+1$ in all results in this 
article). Our treatment of the constraint is also identical to the treatment in Ref.~\onlinecite{tong2016}, 
which discusses new Chern-Simons matrix models which can describe non-Abelian FQH states (our $m$ is equal to their
$k+1$ for their model with $p=1$).}. Finally,
we define ${G^j}_k$ to be the $j,k$ matrix element of the constraint, but divided by a factor of $\hbar$ for convenience,
\beq
	{G^j}_k= i\sum_{A,B,C=1}^{N^2-1}a^{\dg}_A a_B f^{ABC}{(T^C)^j}_k +  \left( \frac{\theta}{\ell_B^2}-1\right) \delta^j_k - b^{\dg}_k b^j \ . \label{eq:constraint-2nd-form}
\eeq

In the quantum theory physical states $|\psi\ran$ will be those states which satisfy 
\beq
	{G^j}_k|\psi\ran= 0\ ,\ \forall\ j,k\ .
\eeq
To understand the form of the physical states $|\psi\ran$ we now analyze the constraint. First set $j=k$ and sum over
all $j$. Then the constraint implies that
\beq
	b^{\dg}_j b^j |\psi\ran= N\left( \frac{\theta}{\ell_B^2}-1\right)|\psi\ran\ .
\eeq
Now we already know that $\theta$ is quantized as an integer, $\theta=\ell_B^2 m$, $m\in \mathbb{Z}$. If we take $m>0$,
then this equation reads as
\beq
	b^{\dg}_j b^j |\psi\ran= N(m-1)|\psi\ran\ . \label{eq:U1-constraint}
\eeq
Thus, we find that the total number of $b^j$ quanta in physical states must be equal to $N(m-1)$. 

Next, we consider the off-diagonal components of the constraint. For this it is convenient to instead consider 
\beq
	G^A := {G^j}_k {(T^A)^k}_j\ ,
\eeq
which is the trace of the product of the constraint matrix (with elements ${G^j}_k$) and a generator $T^A$ of $SU(N)$.
We find that these operators take the form
\beq
	G^A= -i \left(\mathcal{O}_{X}(T^A) + \mathcal{O}_{\Psi}(T^A) \right)\ ,
\eeq
where $\mathcal{O}_{X}(T^A)$ and $\mathcal{O}_{\Psi}(T^A)$ are the quantum operators which generate the
action of the $SU(N)$ generator $T^A$ on the $X^a$ and $\Psi$ variables, respectively. We define these operators and
demonstrate their properties in Appendix~\ref{app:generators}. Thus, the set of constraints
\beq
	G^A|\psi\ran = 0\ ,\ A=1,\dots, N^2-1\ 
\eeq
simply expresses the fact that physical states must be singlets under the total $SU(N)$ action, as originally noted by 
Polychronakos~\cite{P1}. 

To summarize, we find that the constraint in the CSMM breaks up into two separate parts. The first is associated with the
$U(1)$ part of the total $U(N)$ action and states that physical states $|\psi\ran$ obey Eq.~\eqref{eq:U1-constraint}. 
The second part is associated with the $SU(N)$ part of $U(N)$ and states that physical states should be 
singlets under the $SU(N)$ action. Now that we understand the constraint, we can write down a basis of physical
states satisfying this constraint. To this end we introduce the matrix-valued operator\footnote{Perhaps a more precise notation for 
this operator would be 
$A^{\dg}= a^{\dg}_0\otimes \frac{\mathbb{I}}{\sqrt{N}} + \sum_{B=1}^{N^2-1} a^{\dg}_B\otimes T^B$, which expresses
the fact that $A^{\dg}$ acts on the tensor product $\mathcal{H}_Q\otimes\mathcal{H}_N$ of an infinite-dimensional Hilbert 
space $\mathcal{H}_Q$ which arises upon quantization of the model, and an $N$-dimensional vector space $\mathcal{H}_N$ 
on which the classical matrix variables $X^a$ act.}
\beq
	A^{\dg}= a^{\dg}_0 \frac{\mathbb{I}}{\sqrt{N}} + \sum_{B=1}^{N^2-1} a^{\dg}_B T^B
\eeq
with matrix elements
\beq
	{(A^{\dg})^j}_k= a^{\dg}_0 \frac{1}{\sqrt{N}}\delta^j_k + \sum_{B=1}^{N^2-1} a^{\dg}_B {(T^B)^j}_k\ .
\eeq
Then, as was shown by Hellerman and Van Raamsdonk in Ref.~\onlinecite{HVR}, 
one possible basis for all physical states is given by states of the form
\beq
	|\{c_1,\dots,c_N\}\ran= \text{Tr}[(A^{\dg})^N]^{c_N}\cdots\text{Tr}[A^{\dg}]^{c_1}|\psi_0\ran 
\eeq
where each $c_j \in \mathbb{N}$ for $j=1,\dots,N$, and 
\beq
	|\psi_0\ran= \left[\ep^{j_1\cdots j_N}b^{\dg}_{j_1}(b^{\dg}A^{\dg})_{j_2}\cdots(b^{\dg}(A^{\dg})^{N-1})_{j_N}\right]^{(m-1)}|0\ran\ .\label{eq:CSMM-ground-state}
\eeq
Note that all $U(N)$ indices $j,k,$ etc. are contracted in these expressions, and so every operator present is a
singlet under the $SU(N)$ action. The overall power of $m-1$ in $|\psi_0\ran$ is required to satisfy the $U(1)$ part of the
constraint coming from Eq.~\eqref{eq:U1-constraint}.

Since the Hamiltonian of the CSMM just counts the total number of $a_A$ quanta in a state, we find that
$|\psi_0\ran$ is the unique ground state of the CSMM, and that it has an energy
\beqa
	E_0 &=&  \hbar\tomega \left[\frac{N^2}{2} + \frac{1}{2}(m-1)N(N-1)\right] \nnb \\
	&=& \hbar\tomega\left[\frac{1}{2}mN^2 +\left(\frac{1-m}{2}\right)N\right]\ .
\eeqa
The excited states $|\{c_1,\dots,c_N\}\ran$ then have an energy
\beq
	E(\{c_1,\dots,c_N\})= E_0 + \hbar\tomega \sum_{j=1}^N c_j j\ .
\eeq
It follows that the partition function of the CSMM at an inverse temperature $\beta$ is 
just
\beq
	Z= \text{Tr}_Q[e^{-\beta H_{CSMM}}]= q^{ \frac{1}{2}mN^2 +\left(\frac{1-m}{2}\right)N}\prod_{j=1}^N\frac{1}{1-q^j}\ ,
\eeq
where $\text{Tr}_Q[\cdot]$ denotes a trace over the quantum Hilbert space (consisting of physical states obeying the
constraint of the CSMM), and where we defined $q= e^{-\beta \hbar\tomega}$. As $N\to\infty$ the product
$\prod_{j=1}^N\frac{1}{1-q^j}$ becomes the partition function for the oscillator modes of a single chiral boson,
which we know is the edge theory of a Laughlin fractional quantum Hall state.

\subsection{Density of the droplet}

Here we review the calculation of the density of the FQH droplet described by the CSMM in the large $N$ limit. We will
see from this calculation that the CSMM with $\theta=\ell_B^2 m$ corresponds to the  
Laughlin state at filling fraction $\nu=\frac{1}{m}$. We do not find $\nu=\frac{1}{m+1}$ as we
treated the constraint of Eq.~\eqref{eq:constraint} \emph{as is} instead of normal-ordering it as in Polychronakos'
original paper~\cite{P1}. 

We compute the density of the droplet following the reasoning outlined by Polychronakos~\cite{P1}. 
The key is to examine the eigenvalue of the operator 
\beq
	\text{Tr}\left\{ \delta_{ab} X^a X^b \right\}= \sum_{A=0}^{N^2-1}\delta_{ab}x^a_A x^b_A
\eeq 
in the ground state $|\psi_0\ran$ of the CSMM (the trace here is a matrix trace). Since this operator is proportional
to $H_{CSMM}$ we have $\text{Tr}\left\{ \delta_{ab} X^a X^b \right\}|\psi_0\ran= R^2|\psi_0\ran$ where
the eigenvalue $R^2$ is given by
\beq
	R^2= 2\ell_B^2\left(m\frac{N(N-1)}{2} +\frac{N}{2} \right)\ . \label{eq:droplet-radial-positions}
\eeq
We interpret this eigenvalue as a sum
of contributions from $N$ different particles at different radial positions by writing it as
\beq
	R^2= \sum_{j=1}^{N}R^2_j\ ,
\eeq
where 
\beq
	R^2_j= 2\ell_B^2\left(m (j-1) + \frac{1}{2}\right)\ .
\eeq
Indeed, the $R^2_j$ can be thought of as the eigenvalues of the classical matrix $\delta_{ab}X^a X^b$, since the
operator $R^2$ is equal to the trace of this matrix.
Thus, we think of the ground state of the droplet as containing $N$ particles at definite radial positions $R_j$ but with
complete uncertainty in their angular position. 
%Note also that due to the offset of $\frac{1}{2}$, the particle closest to the
%origin is at radius $\ell_B$ instead of at zero, so no particle has a definite position in space. 
In addition, since $R^2_j$ is linear in $j$, the area $\pi (R^2_j- R^2_{j-1})=2\pi\ell_B^2 m$ of the annulus 
between 
consecutive particles is independent of $j$. This implies that the particles are distributed uniformly, i.e., the density is a 
constant within the droplet.

The size of the droplet is given by
the largest value of $R^2_j$, which is
\beq
	R^2_{N} = 2\ell_B^2\left(m(N-1)+\frac{1}{2}\right)\approx 2\ell_B^2 m N
\eeq
for large $N$. Then at large $N$ we compute the density as being that of $N$ particles evenly spread out over a disk
of radius $R^2_{N}\approx 2\ell_B^2 m N$, and so 
\beq
	\rho_0 = \frac{N}{\pi R^2_{N}} \approx \frac{1}{2\pi \ell_B^2 m}\ ,
\eeq
which is exactly the density of the Laughlin state with filling fraction $\nu=\frac{1}{m}$ (in the limit of a large number $N$
of electrons).

\section{Hall viscosity of the CSMM}
\label{sec:CSMM-viscosity}

We now compute the Hall viscosity in the CSMM following the calculation of Park and Haldane~\cite{park-haldane} (which
we reviewed in Sec.~\ref{sec:GC-hall-visc}). We find that the Hall viscosity tensor contains only a \emph{single} contribution,
and that this contribution is equal to the guiding center Hall viscosity of the Laughlin state. In other words, the CSMM
lacks the Landau orbit contribution to the Hall viscosity, but does contain the (physically important) guiding center contribution.

To compute the Hall viscosity in this system we recall that in the fluid interpretation of the NCCS theory
and the CSMM (which we reviewed at the end of Sec.~\ref{sec:NCCS}), 
the variables $X^a$ represent a noncommutative analogue of fluid coordinates in 
a Lagrange description of a fluid~\cite{susskind,P1,P2}. In the case of the CSMM, this is a finite droplet of noncommutative 
fluid. Thus, to compute the Hall viscosity we first need to identify the quantum operators $\mathsf{\Lambda}^{ab}$ which 
generate APDs (or strains) of the noncommutative fluid coordinates $X^a$. 
Since we expand the noncommutative coordinates in terms of
the scalar variables $x^a_A$, $A=0,\dots,N^2-1$, we can instead search for operators which implement APDs 
of these variables. These operators will then automatically implement the correct transformations of the
$X^a$ coordinates, as the operators do not act on the matrix indices of the $X^a$ variables. 

Since the commutation relations of the variables $x^a_A$ are identical to the commutation relations of the 
guiding center coordinates in the quantum Hall problem, we immediately see that the desired operators are given by
\beq
	\mathsf{\Lambda}^{ab}=\frac{1}{4\ell_B^2}\sum_{A=0}^{N^2-1} \{ x^a_A, x^b_A\}\ .
	%\Lambda^{ab}= \frac{1}{4\ell_B^2} \{ {(X^a)^j}_k, {(X^b)^k}_j  \}\ .
\eeq
These operators obey the same algebra as in Eq.~\eqref{eq:APD-algebra-gc}. It follows that the unitary operators which 
implement the APDs are $U(\al)= e^{i\al_{ab}\mathsf{\Lambda}^{ab}}$, with $\al_{ab}$ a constant
symmetric matrix. To first order in $\al_{ab}$ we have (for all $A=0,\dots,N^2-1$)
\beq
	U(\al) x^a_A U(\al)^{\dg}= x^a_A + \ep^{ab}\al_{bc}x^c_A+\cdots
\eeq
which implies (for all $j,k$)
\beq
	U(\al) {(X^a)^j}_k U(\al)^{\dg}= {(X^a)^j}_k + \ep^{ab}\al_{bc}{(X^c)^j}_k+\cdots
\eeq

It is important to note that the APD generators $\mathsf{\Lambda}^{ab}$ act only on the physical position indices $a$ of the
variables $X^a$. There is no action at all on the $U(N)$ indices $j,k$ of the matrix elements $ {(X^a)^j}_k$. Thus,
the generators $\mathsf{\Lambda}^{ab}$ act identically on all matrix elements of $X^a$, and so they are indeed the correct 
quantum generators of APDs of the noncommutative fluid coordinates $X^a$ (which we recall are actually 
$N\times N$ Hermitian matrices in the classical theory).

Now we want to compute the Hall viscosity in the ground state $|\psi_0\ran$ of the CSMM. We compute
this using a Kubo formula approach similar to that of Ref.~\onlinecite{bradlyn2012}. 
We present the Kubo formula calculation of the Hall viscosity in Appendix~\ref{app:Kubo}. Our result is that
the Hall viscosity tensor in this model takes the form ($A$ is the area of the droplet)
\beq
	\eta^{abcd}_{\text{{\tiny{CSMM}}}}= \frac{i\hbar}{A} \lan \psi_0| [\mathsf{\Lambda}^{ab},\mathsf{\Lambda}^{cd}]|\psi_0\ran\ .
\eeq 
We note that the tensor $\eta^{abcd}_{\text{{\tiny{CSMM}}}}$ contains only a single contribution, as opposed to the 
two separate terms (guiding center and Landau orbit contributions) appearing in the discussion of the Hall viscosity tensor from 
Sec.~\ref{sec:GC-hall-visc}.
Note that in deriving this result it was crucial
that the CSMM has a unique ground state and a finite energy gap set by the frequency $\tomega$ of the harmonic trap. 

Due to the commutation relations of the generators $\mathsf{\Lambda}^{ab}$ (which are the same as 
Eq.~\eqref{eq:APD-algebra-gc}), the four index tensor $\eta^{abcd}_{\text{{\tiny{CSMM}}}}$
can again be expressed in terms of a symmetric two-index tensor
\beq
	\eta^{ab}_{\text{{\tiny{CSMM}}}} = -\frac{\hbar}{A}\lan \psi_0|\mathsf{\Lambda}^{ab}|\psi_0\ran \ .
\eeq
Therefore, to compute the Hall viscosity tensor of the CSMM, we just need to compute the expectation values 
$\lan \psi_0|\mathsf{\Lambda}^{ab}|\psi_0\ran$. To compute these we first note that the CSMM Hamiltonian can be
written as
\beq
	H_{CSMM}= \hbar\tomega \delta_{ab}\mathsf{\Lambda}^{ab}= \hbar \tomega (\mathsf{\Lambda}^{11}+\mathsf{\Lambda}^{22})\ .
\eeq
From this we can already deduce that
\beq
	\lan \psi_0|\delta_{ab}\mathsf{\Lambda}^{ab}|\psi_0\ran= \frac{E_0}{\hbar\tomega}= \frac{1}{2}mN^2 +\left(\frac{1-m}{2}\right)N\ . \label{eq:trace-strain}
\eeq 
We can go further and compute the individual expectation values of $\mathsf{\Lambda}^{11}$ and $\mathsf{\Lambda}^{22}$ 
by deriving a Virial theorem for the CSMM. To derive this theorem consider the operator
\beq
	Q= \sum_{A=0}^{N^2-1} x^1_A x^2_A\ .
\eeq
A short computation shows that
\beq
	[Q,\delta_{ab}\mathsf{\Lambda}^{ab}]= 2i\ell_B^2(-\mathsf{\Lambda}^{11}+\mathsf{\Lambda}^{22})\ .
\eeq
If we take the expectation value of this equation in the state $|\psi_0\ran$ (or any eigenstate of 
$\delta_{ab}\mathsf{\Lambda}^{ab}$), then we find that
\beq
	\lan \psi_0|\mathsf{\Lambda}^{11}|\psi_0\ran= \lan \psi_0|\mathsf{\Lambda}^{22}|\psi_0\ran\ .
\eeq
Combining this result with Eq.~\eqref{eq:trace-strain} gives the result that
\beq
	\lan \psi_0|\mathsf{\Lambda}^{11}|\psi_0\ran= \lan \psi_0|\mathsf{\Lambda}^{22}|\psi_0\ran= \frac{1}{2}\left[\frac{1}{2}mN^2 +\left(\frac{1-m}{2}\right)N \right]\ .
\eeq

Finally, it remains to compute the expectation value of the off-diagonal generator 
$\mathsf{\Lambda}^{12}=\mathsf{\Lambda}^{21}$. 
In terms of the oscillator variables $a_A$ and $a^{\dg}_A$ this operator takes the form
\beq
	\mathsf{\Lambda}^{12}= \frac{1}{4i}\sum_{A=0}^{N^2-1}\left( a_A a_A - a^{\dg}_A a^{\dg}_A \right)\ .
\eeq
Now all eigenstates of $H_{CSMM}$ are eigenstates of the total number operator for the $a_A$ oscillators. Since
$\mathsf{\Lambda}^{12}$ clearly does not commute with the total number operator, we immediately conclude that 
the expectation value of $\mathsf{\Lambda}^{12}$ in any eigenstate of $H_{CSMM}$ is zero. 

Therefore our final result for the expectation value of the APD generators 
$\mathsf{\Lambda}^{ab}$ in the CSMM ground state is
\beq
	\lan \psi_0|\mathsf{\Lambda}^{ab}|\psi_0\ran=  \frac{1}{2}\left[\frac{1}{2}mN^2 +\left(\frac{1-m}{2}\right)N \right]\delta^{ab}\ .
\eeq
This means that we can write $\eta^{ab}_{\text{{\tiny{CSMM}}}} = \eta_{\text{{\tiny{CSMM}}}}\delta^{ab}$, where
the coefficient $\eta_{\text{{\tiny{CSMM}}}}$ of Hall viscosity in this model is equal to 
\beq
	\eta_{\text{{\tiny{CSMM}}}} = -\frac{\hbar}{A}\frac{1}{2}\left[\frac{1}{2}m N^2 + \left( \frac{1-m}{2} \right)N\right]\ .
\eeq
Now since $A= \pi R_N^2\approx 2\pi \ell_B^2 m N$ for the CSMM at large $N$, this exactly matches the result (before 
regularization) for the \emph{guiding center Hall viscosity} $\eta_H$ of the  $\nu=\frac{1}{m}$ 
Laughlin state. The Landau orbit contribution $\td{\eta}_H$ is absent in the CSMM. 
Finally, as was the case for the ordinary Laughlin state, 
this result can be regularized by subtracting off 
the extensive term in $\eta_{\text{{\tiny{CSMM}}}}$ 
(or the superextensive term in $\lan \psi_0|\mathsf{\Lambda}^{ab}|\psi_0\ran$).
We discuss a fluid interpretation of this regularization of the Hall viscosity later in Sec.~\ref{sec:reg}.

\section{Hall conductance of the CSMM in a non-uniform electric field}
\label{sec:hall-conductance}

In this section we study the Hall conductance of the CSMM when it is subjected to a non-uniform electric field. Our motivation
for studying this setup is the well-known result of Hoyos and Son which shows that in a quantum Hall state 
the Hall conductance $\sigma_{H}(\mb{k})$ at finite wave vector $\mb{k}$ has a universal contribution at order $k^2$ 
($k^2= \delta^{ab}k_a k_b$) which is related to the Hall viscosity~\cite{hoyos-son} (see also Ref.~\onlinecite{bradlyn2012}
for a Kubo formula approach to this relation). 
We find a similar contribution in the CSMM, but depending only on the guiding center Hall viscosity as opposed to the
full Hall viscosity. Again, this is not surprising as we only expect the CSMM to describe the dynamics
of the guiding center degrees of freedom in a FQH state.

In this section we first review the result of Ref.~\onlinecite{hoyos-son} on the Hall conductance at finite wave vector. We
then warm up by calculating the Hall conductance of the CSMM subjected to a \emph{uniform} electric field. The reason for
this is that there are several subtle points associated with the computation of the Hall conductance in the CSMM that we
want to explain clearly. Finally, we compute the Hall conductance of the CSMM in a 
non-uniform electric field, where we find a result which resembles the result of Hoyos and Son~\cite{hoyos-son}, but 
with the full Hall viscosity replaced by the guiding center Hall viscosity. We note here
that the Hall conductance of the NCCS theory in a uniform electric
field was computed previously in Refs.~\onlinecite{hansson2001,hansson2003} at the classical level by solving the 
equations of motion for the NCCS theory in a uniform electric field. We therefore emphasize that our treatment in this section 
deals directly with the quantized CSMM theory as opposed to the classical NCCS theory. 

\subsection{The result of Hoyos and Son}

We start by reviewing the result of Ref.~\onlinecite{hoyos-son}. Consider a quantum Hall system in a non-uniform
electric field $\mb{E}= (E(\mb{x}),0)$ pointing in the $x^1$ direction, and where the
spatial dependence is only on the $x^1$ coordinate, so that $\pd_2 E(\mb{x})=0$. 
The Hall conductance $\sigma_{H}(\mb{k})$ at finite wave vector is defined by the relation
\beq
	j^2(\mb{k})= \sigma_{H}(\mb{k}) E(\mb{k})\ ,
\eeq
where $j^2(\mb{k})$ is the Fourier transform of the charge current in the $x^2$ direction, and $E(\mb{k})$ is the
Fourier transform of $E(\mb{x})$. The result of  Ref.~\onlinecite{hoyos-son} is that (recall that $E(\mb{x})$ is a function 
of only $x^1$)
\beq
	\frac{\sigma_H(\mb{k})}{\sigma_H(\mb{0})}= 1 + C_2 (k_1 \ell_B)^2+\cdots\ ,
\eeq
where the Hall conductance at zero wave vector is simply ($\nu$ is the filling fraction)
\beq
	\sigma_H(\mb{0})= \nu \frac{e^2}{h}\ .
\eeq
The coefficient $C_2$ is given by
\beq
	C_2=\frac{\eta_{tot}}{\hbar\rho_0} - \frac{2\pi}{\nu}\frac{\ell_B^2}{\hbar\omega_c}B^2\mathcal{E}''(B)\ , 
\label{eq:hoyos-son}
\eeq
where $\eta_{tot}$ denotes the full Hall viscosity of the quantum Hall state (as opposed to just the guiding center
part), $\mathcal{E}(B)$ is the energy density of the quantum Hall state viewed as a function of the external field
$B$, and $\mathcal{E}''(B)$ denotes the second derivative of $\mathcal{E}(B)$ with respect to $B$. In addition,
$\rho_0$ denotes the number density of the quantum Hall state, and $\omega_c=\frac{eB}{M}$ is the cyclotron frequency,
where $M$ is the mass of the particles making up the quantum Hall state. As an example, for a 
quantum Hall state consisting of $N$ electrons in the lowest Landau level and occupying an area $A$, we have
$\mathcal{E}(B)= \frac{\hbar \omega_c}{2}\frac{N}{A}= \frac{\hbar \omega_c}{2}\rho_0$, and for a Laughlin 
$\nu=\frac{1}{m}$ state this gives $\mathcal{E}(B)= \frac{\hbar\omega_c}{4\pi \ell_B^2 m}$. 

In the context of the CSMM, the quantity that we actually compute is the current at the location of the center of mass
of the droplet (we explain the reason for this in the next subsection). 
Therefore we need to Fourier transform the result of Hoyos and Son back to real space in order to compare
with our calculation in the CSMM later in this section. In real space we find that
\beq
	j^2(\mb{x})= \nu \frac{e^2}{h}\left( E(\mb{x}) - C_2 \ell_B^2 \pd^2_1E(\mb{x}) + \dots \right)\ .
\eeq
In particular, at the origin $\mb{x}=\mb{0}$ (where the center of mass of a uniform droplet would be located) we have
\beq
	j^2(\mb{x}=\mb{0})= \nu \frac{e^2}{h}\left( E^{(0)} - C_2 \ell_B^2 E^{(2)} + \dots \right)\ , \label{eq:current-real}
\eeq
where $E^{(0)}$ and $E^{(2)}$ are the coefficients in the Taylor series expansion of $E(\mb{x})$ about the
origin,
\beq
	E(\mb{x})= E^{(0)} + E^{(1)} x^1 + \frac{1}{2!} E^{(2)} (x^1)^2 + \dots\ ,
\eeq
and where we again remind the reader that we assumed that $E(\mb{x})$ has no $x^2$ dependence. 

\subsection{Uniform electric field}

We now compute the Hall conductance of the CSMM in a uniform electric field. Our reason for treating this simple
case first is to highlight a few subtleties in the calculation of the Hall conductance of the CSMM. The first subtlety is 
associated with the fact that one cannot resolve individual points in space in the CSMM, since the spatial coordinates
are actually the noncommuting matrices $X^1$ and $X^2$. However, in the CSMM one can still define a notion of the
center of mass coordinate of the FQH droplet, and the expectation value of this center of mass coordinate can be
computed in any state $|\psi\ran$ of the quantized CSMM. We define the center of mass coordinates $X^a_{\text{{\tiny{COM}}}}$
as
\beq
	X^a_{\text{{\tiny{COM}}}}= \frac{1}{N}\text{Tr}\{X^a\} = \frac{x^a_0}{\sqrt{N}}\ ,
\eeq
where in the second equality we evaluated the trace and found that $X^a_{\text{{\tiny{COM}}}}$ is proportional to the variable
$x^a_0$ introduced in Eq.~\eqref{eq:expansion} of Sec.~\ref{sec:CSMM}.
To motivate this definition we simply note that if the $X^a$ were diagonal matrices, then their diagonal elements could 
be interpreted as the positions of $N$ particles, and then $\frac{1}{N}\text{Tr}\{X^a\}$ would agree with the usual
definition of the center of mass coordinate of $N$ particles (assuming all particles have equal masses).

Our strategy to compute the Hall conductance in the CSMM is to compute the drift velocity $\mb{v}_{drift}$ of the
center of mass coordinate when the system is placed in an electric field $\mb{E}$. We can then use the fact that the
CSMM describes a droplet of particles with charge $-e$ and density $\rho_0=\frac{1}{2\pi \ell_B^2 m}$ (computed in 
Sec.~\ref{sec:CSMM}) to 
compute the charge current $\mb{j}_{\text{{\tiny{COM}}}}$ at the center of mass as
\beq
	\mb{j}_{\text{{\tiny{COM}}}} = -e\rho_0 \mb{v}_{drift}\ .
\eeq
The result can then be compared with the result of Hoyos and Son for the current at the origin (location of the center of mass)
as expressed in Eq.~\eqref{eq:current-real}.

Next, we need to discuss the issue of how to couple the CSMM to an external electric field. This can be done using
the fluid interpretation of this theory from Sec.~\ref{sec:NCCS}. First, recall from Sec.~\ref{sec:NCCS} that
an ordinary charged fluid on commutative flat space $\mathbb{R}^2$ can be coupled to a background electromagnetic
field by including vector and scalar potentials $\mathcal{A}_a(t,\mb{X})$ and $\vphi(t,\mb{X})$, respectively, in the action for 
the Lagrange description of this fluid, Eq.~\eqref{eq:charged-fluid}. In our case we are only interested in
adding a scalar potential $\vphi(t,\mb{X})$ for the external electric field. Using the fluid interpretation 
we can incorporate this potential into the NCCS theory by adding a term to the NCCS action of the form
\beq
	S_{EM}= e\int_0^T dt\ \text{Tr}_{\mathcal{H}_F}\left\{ \hat{\vphi}(\hat{\mb{X}},t)\right\}\ ,
\eeq
where the operator $\hat{\vphi}(\hat{\mb{X}},t)$ is the operator representing the scalar potential for the
external electromagnetic field (and recall that the charge of the particles is $q=-e$).

In defining the operator $\hat{\vphi}(\hat{\mb{X}},t)$ we encounter an ordering ambiguity. For example
if the scalar potential for the electric field configuration on a commutative space is
$\vphi(t,\mb{X})= X^1X^2$, then we could define $\hat{\vphi}(\hat{\mb{X}},t)= \hat{X}^1\hat{X}^2$, 
$\hat{\vphi}(\hat{\mb{X}},t)= \hat{X}^2\hat{X}^1$, or the symmetric Weyl ordering 
$\hat{\vphi}(\hat{\mb{X}},t)= \frac{1}{2}\left(\hat{X}^1\hat{X}^2+\hat{X}^2\hat{X}^1\right)$, for example. 
We choose
to use Weyl ordering since this is consistent with our use of Weyl ordering to go between star product
and operator formulations of noncommutative field theory (recall Eq.~\eqref{eq:Weyl-ordered}), however, in
the examples of this section we do not actually encounter this ordering ambiguity. Weyl-ordering for the
external field was also adopted by the authors of Ref.~\onlinecite{hansson2001}, 
who also considered the NCCS theory in the presence of external fields. 

Finally, to couple the CSMM to the
external electromagnetic field we use the same action $S_{EM}$ as above but replace the operators $\hat{X}^a$
on the infinite-dimensional space $\mathcal{H}_F$ with the finite $N\times N$ matrix variables of the CSMM. From this
action we can then read off the new Hamiltonian for the CSMM coupled to the external electric field.

There is one more subtlety with the calculation of the Hall conductance of the CSMM that we need to address before
we can proceed. The issue is that the parabolic potential in the CSMM competes with the applied electric field to determine
the long time behavior of the CSMM in the presence of the electric field. This is best illustrated for the case of the 
CSMM in a constant electric field $E^{(0)}$ pointing in the $x^1$ direction. The Hamiltonian describing this system is
\beqa
	H' &=& H_{CSMM} + e E^{(0)}\text{Tr}\{X^1\} \nnb \\
	&=& H_{CSMM} + e E^{(0)}N X^1_{\text{{\tiny{COM}}}}\ , \label{eq:CSMM-constant-E}
\eeqa
and where the trace is a classical matrix trace. To derive this Hamiltonian we used the
fluid interpretation of the CSMM theory and incorporated a scalar potential $\vphi(t,\mb{X})= -E^{(0)}X^1$ to describe
the coupling to a constant electric field in the $x^1$ direction. 
This Hamiltonian can be immediately diagonalized by noting that
\beq
	H' = T(\mb{R})H_{CSMM} T(\mb{R})^{\dg} - \frac{e N (E^{(0)})^2 }{2 B\tomega}\ ,
\eeq
where $T(\mb{R})$ is a unitary translation operator\footnote{We have
$[X^a_{\text{{\tiny{COM}}}},X^b_{\text{{\tiny{COM}}}}]= \frac{i\ell_B^2}{N}\ep^{ab}$ and 
$ T(\mb{R})X^a_{\text{{\tiny{COM}}}} T(\mb{R})^{\dg}= X^a_{\text{{\tiny{COM}}}}+ R^a$.} (similar to a magnetic 
translation) of the form
\beq
	T(\mb{R})= \text{exp}\left\{-\frac{i \ep_{ab} N X^a_{\text{{\tiny{COM}}}} R^b}{\ell_B^2}\right\}\ ,
\eeq
and where in this case
\beq
	\mb{R}= \left(\frac{E^{(0)}}{B\tomega},0\right)\ .
\eeq
The ground state of this Hamiltonian is $|\psi_0'\ran= T(\mb{R})|\psi_0\ran$ and represents a stationary state
with $\lan \psi_0'|X^1_{\text{{\tiny{COM}}}}|\psi_0'\ran = -\frac{E^{(0)}}{B\tomega}$ and 
$\lan \psi_0'|X^2_{\text{{\tiny{COM}}}}|\psi_0'\ran=0$, which corresponds to the equilibrium position in the total potential
\beq
	V= \frac{eBN\tomega}{2}\delta_{ab}X^a_{\text{{\tiny{COM}}}}X^b_{\text{{\tiny{COM}}}} + e E^{(0)}N X^1_{\text{{\tiny{COM}}}}
\eeq
felt by the center of mass. 

We see that if we simply diagonalize the Hamiltonian $H'$ for the CSMM in the 
presence of the external field, we find
no time dependence and, in the ground state, the center of mass of the droplet just sits at its equilibrium position 
$(-\frac{E^{(0)}}{B\tomega},0)$ under the influence of the combined forces of the parabolic potential and the 
applied electric field. 

To compute the Hall conductance of this model we instead need to consider a non-equilibrium situation in which we start
with the system in the ground state $|\psi_0\ran$ of the unperturbed CSMM (which we will now assume is properly 
normalized) and then suddenly turn on the 
electric field. We then study the time evolution of the center of mass coordinate at small times 
$t\ll \frac{1}{\tomega}$, where $\frac{1}{\tomega}$ is the time scale set by the parabolic potential. Therefore
we consider the ``quantum quench" problem in which the state of the system at time $t$ is given by
\beq
	|\psi(t)\ran= e^{-i\frac{H't}{\hbar}}|\psi_0\ran\ ,
\eeq
where $|\psi_0\ran$ is the ground state of the unperturbed CSMM Hamiltonian $H_{CSMM}$, and $H'$ is the perturbed
CSMM Hamiltonian including the applied electric field. We then compute
\begin{align}
	\lan \psi(t)|X^a_{\text{{\tiny{COM}}}}|\psi(t)&\ran = \lan \psi_0|X^a_{\text{{\tiny{COM}}}}|\psi_0\ran \nnb \\
 +\ \frac{it}{\hbar}&\lan\psi_0|[H',X^a_{\text{{\tiny{COM}}}}] |\psi_0\ran + \dots
\end{align}
and identify the drift velocity $\mb{v}_{drift}$ of the center of mass with the term linear in $t$ in this expansion,
\beq
	v^a_{drift}= \frac{i}{\hbar}\lan\psi_0|[H',X^a_{\text{{\tiny{COM}}}}] |\psi_0\ran\ . \label{eq:v-drift}
\eeq

We now consider the case of a uniform electric field $E^{(0)}$ pointing in the $x^1$ direction so that $H'$ takes the form 
shown in Eq.~\eqref{eq:CSMM-constant-E}. In this case the drift velocity evaluates to 
\beq
	\mb{v}_{drift}= \left(0,-\frac{E^{(0)}}{B} \right)\ .
\eeq
Then the non-zero part of the charge current at the center of mass of the droplet, at times $t\ll \frac{1}{\tomega}$, is
\beqa
	j^2_{\text{{\tiny{COM}}}} &=& e\rho_0\frac{E^{(0)}}{B} \nnb \\
	&=& \nu \frac{e^2}{h} E^{(0)}\ ,
\eeqa
with $\nu=\frac{1}{m}$, and where we used $\rho_0=\frac{1}{2\pi \ell_B^2 m}$. Therefore we find that the
Hall conductance of the CSMM with $\theta=\ell_B^2 m$ is given by
\beq
	\sigma_H= \frac{1}{m}\frac{e^2}{h}\ ,
\eeq
exactly as in the $\nu=\frac{1}{m}$ Laughlin state.

For the case of a uniform electric field we can actually go further and compute the full time dependence of the center of mass
coordinate. We find that
\beqa
	\lan \psi(t)|X^1_{\text{{\tiny{COM}}}}|\psi(t)\ran &=& \frac{E^{(0)}}{B\tomega}\left( -1+\cos(\tomega t) \right) \\
	\lan \psi(t)|X^2_{\text{{\tiny{COM}}}}|\psi(t)\ran &=& -\frac{E^{(0)}}{B\tomega}\sin(\tomega t)\ .
\eeqa
We see that the center of mass moves in a large circle about its equilibrium position $(-\frac{E^{(0)}}{B\tomega},0)$, but
that at early times $t\ll \frac{1}{\tomega}$ the droplet drifts in the $x^2$ direction with velocity vector
$\mb{v}_{drift}= \left(0,-\frac{E^{(0)}}{B} \right)$.

% In fact, we can actually take the $\tomega\to 0$ limit to find
%$\lim_{\tomega\to 0} \lan \psi(t)|X^1_{\text{{\tiny{COM}}}}|\psi(t)\ran =0$ and 
%$\lim_{\tomega\to 0} \lan \psi(t)|X^2_{\text{{\tiny{COM}}}}|\psi(t)\ran = -\frac{E^{(0)}}{B}t$, which represents a droplet 
%moving straight in the negative $x^2$ direction. 

\subsection{Non-uniform electric field}

We now compute the Hall conductance of the CSMM in a non-uniform electric field. We consider an electric field which
points in the $x^1$ direction, and which depends only on the $x^1$ coordinate. Since we are interested in contributions
to the current which depend on the second derivative of the electric field, it is sufficient to consider an electric field which
depends at most quadratically on the $x^1$ coordinate. Thus, for an ordinary classical charged fluid
described by the action of Eq.~\eqref{eq:charged-fluid}, we would add a scalar potential of the form
\beq
	\vphi(t,\mb{X})= -E^{(0)}X^1 - \frac{1}{2}E^{(1)}(X^1)^2- \frac{1}{3!}E^{(2)}(X^1)^3\ ,
\eeq
which corresponds, after computing minus the spatial gradient, to an electric field $\mb{E}=(E(\mb{X}),0)$ with
\beq
	E(\mb{X})= E^{(0)} + E^{(1)}X^1 + \frac{1}{2}E^{(2)}(X^1)^2\ .
\eeq
The coefficients $E^{(j)}$, $j=0,1,2$ in this expression (which are all fixed real numbers) can be understood as the 
coefficients in the Taylor expansion of $E(\mb{X})$ about the origin. 

This form of the scalar potential for the ordinary classical fluid, combined with the considerations from earlier in this section
on how to couple the CSMM to external fields, leads to a Hamiltonian 
\beq
	H'= H_{CSMM} + H_1
\eeq 
with
\beq
	H_1= e\text{Tr}\left\{ E^{(0)}X^1 + \frac{1}{2}E^{(1)}(X^1)^2+ \frac{1}{3!}E^{(2)}(X^1)^3 \right\}\ ,
\eeq
where the trace denotes a matrix trace. This Hamiltonian then describes the CSMM in the presence of a non-uniform 
electric field in the $x^1$ direction. 
To compute the Hall conductance we again consider a time-dependent problem where the state at 
time $t$ is given by $|\psi(t)\ran= e^{-i\frac{H't}{\hbar}}|\psi_0\ran$ with $|\psi_0\ran$ the ground state of
$H_{CSMM}$. The drift velocity is again given by Eq.~\eqref{eq:v-drift} and since 
$\lan \psi_0|[H_{CSMM},X^a_{\text{{\tiny{COM}}}}]|\psi_0\ran=0$ (since $|\psi_0\ran$ is an eigenstate of $H_{CSMM}$), 
this reduces to
\beq
	v^a_{drift}= \frac{i}{\hbar}\lan\psi_0|[H_1,X^a_{\text{{\tiny{COM}}}}] |\psi_0\ran\ . 
\eeq
It remains to actually compute the matrix element $\lan\psi_0|[H_1,X^a_{\text{{\tiny{COM}}}}] |\psi_0\ran$. 

To compute this matrix element we first note that we already know the answer for the term in $H_1$ proportional to 
$E^{(0)}$ from the previous subsection. 
Next, we can immediately see that the term proportional to $E^{(1)}$ will vanish since the commutator
of $\text{Tr}\{(X^1)^2\}$ with $X^a_{\text{{\tiny{COM}}}}$ is linear in the center of mass coordinate and we know
that $\lan\psi_0|X^a_{\text{{\tiny{COM}}}} |\psi_0\ran=0$ in the unperturbed ground state of the CSMM. To handle the term 
proportional to $E^{(2)}$ we use Eq.~\eqref{eq:expansion} to find that
\begin{align}
	\text{Tr}\{(X^1)^3\} = &\frac{(x^1_0)^3}{\sqrt{N}}+\frac{3}{\sqrt{N}}x^1_0\sum_{A=1}^{N^2-1}x^1_A x^1_A \nnb \\ +& \sum_{A,B,C=1}^{N^2-1}x^1_A x^1_B x^1_C \text{Tr}\{T^A T^B T^C\}\ .
\end{align}
Then we have $[\text{Tr}\{(X^1)^3\},X^1_{\text{{\tiny{COM}}}}]=0$ and
\beq
	[\text{Tr}\{(X^1)^3\},X^2_{\text{{\tiny{COM}}}}]= \frac{3 i\ell_B^2}{N}\sum_{A=0}^{N^2-1}x^1_A x^1_A \ .
\eeq
We find that $v^1_{drift}=0$, while 
\beqa
	v^2_{drift}&=& -\frac{E^{(0)}}{B} +\frac{i}{\hbar}\left(e\frac{E^{(2)}}{3!}\right)\frac{3 i\ell_B^2}{N}\lan \psi_0|\sum_{A=0}^{N^2-1}x^1_A x^1_A|\psi_0\ran \nnb \\
	&=& -\frac{E^{(0)}}{B} - \frac{e E^{(2)} \ell_B^4}{\hbar N}\lan \psi_0|\mathsf{\Lambda}^{11}|\psi_0\ran \nnb \\
	&=& -\frac{E^{(0)}}{B} + \frac{E^{(2)}\ell_B^2}{B}\frac{\eta_{\text{{\tiny{CSMM}}}}}{\hbar\rho_0}\ ,
\eeqa
where we used the fact that $\lan \psi_0|\mathsf{\Lambda}^{11}|\psi_0\ran= -\frac{A}{\hbar}\eta_{\text{{\tiny{CSMM}}}}$ 
and $\rho_0=\frac{N}{A}$.
If we now compute $j^2_{\text{{\tiny{COM}}}}= -e\rho_0 v^2_{drift}$ then we find that
\beqa
	j^2_{\text{{\tiny{COM}}}} &=& \nu\frac{e^2}{h}\left( E^{(0)} - E^{(2)}\ell_B^2 \frac{\eta_{\text{{\tiny{CSMM}}}}}{\hbar\rho_0} \right) \nnb \\
	&=& \nu\frac{e^2}{h}\left( E^{(0)} - E^{(2)}\ell_B^2 \frac{\eta_H}{\hbar\rho_0} \right)\ ,
\eeqa
where the second line follows from the fact that $\eta_{\text{{\tiny{CSMM}}}}=\eta_H$, where 
$\eta_H$ was the guiding center Hall viscosity for the Laughlin state.
Finally, we should regularize this expression to obtain a finite answer for the current in the $N\to\infty$ limit. This 
just amounts to the replacement $\eta_H \to \eta_{H,reg}$ in the final expression (we discuss the physical
interpretation of this regularization in Sec.~\ref{sec:reg}). Therefore our final expression for the center of mass
current in a non-uniform electric field is
\beq
	j^2_{\text{{\tiny{COM}}}}= \nu\frac{e^2}{h}\left( E^{(0)} - E^{(2)}\ell_B^2 \frac{\eta_{H,reg}}{\hbar\rho_0} \right)\ . \label{eq:CSMM-current-NU}
\eeq
Eq.~\eqref{eq:CSMM-current-NU} is the main result of this section. 

It is interesting to compare Eq.~\eqref{eq:CSMM-current-NU} with the result of Hoyos and Son, Eq.~\eqref{eq:current-real},
where the coefficient $C_2$ was given in Eq.~\eqref{eq:hoyos-son}. We see that the CSMM result contains a contribution
like the first term in $C_2$, but with the total Hall viscosity $\eta_{tot}$ replaced with the guiding center Hall viscosity
$\eta_{H,reg}$. As we remarked earlier, this makes sense because we only expect the CSMM to describe the dynamics
of the guiding center degrees of freedom in the quantum Hall problem. We also find that the CSMM result does not contain
any contribution resembling the second term in $C_2$ which is proportional to $\mathcal{E}''(B)$. This is also not 
surprising since the CSMM itself does not contain any information about the energy associated with electrons filling
a Landau level. 
Indeed, we can see from the fluid interpretation of the NCCS theory from Sec.~\ref{sec:NCCS}
that the NCCS theory (and therefore the CSMM theory which is a regularization of it), is obtained by sending the
energy scale $\hbar \omega_c$ to infinity. Therefore we find that the CSMM accurately captures the
\emph{guiding center contribution} to the response of a FQH state to a non-uniform electric field.

\section{$N\to\infty$ limit, regularization of the Hall viscosity, and fluid interpretation}
\label{sec:reg}

In Ref.~\onlinecite{park-haldane} Park and Haldane argued that one should regularize the guiding center Hall viscosity
by subtracting the extensive term in 
$\eta_{H}= -\frac{\hbar}{A}\frac{1}{2}\left[\frac{1}{2}m N^2 + \left( \frac{1-m}{2} \right)N\right]$, which amounts
to subtracting the term $\frac{1}{2}m N^2$ from  
\beq
	\frac{1}{2}m N^2 + \left( \frac{1-m}{2} \right)N \ .
\eeq
In the quantum Hall problem this regularization (or something similar to it) is necessary to obtain a finite value for 
the guiding center Hall viscosity in the thermodynamic limit $N\to\infty$. 
%In the context of the CSMM there is another
%reason for studying the $N\to\infty$ limit: understanding this limit is necessary for a precise definition of the quantized
%NCCS theory. We will not be able to explore this issue completely in this paper, but we make a few remarks on it in 
%the Conclusion.

In this section we give an interpretation of this regularization scheme
in the context of the fluid interpretation (reviewed in the last subsection of Sec.~\ref{sec:NCCS}) of the NCCS theory and 
CSMM. Our starting point is to note that the expectation value $\lan\psi_0|\mathsf{\Lambda}^{ab}|\psi_0\ran$
in the CSMM is actually proportional to the total angular momentum of the state $|\psi_0\ran$. 
The fact that the Hall viscosity is related to angular momentum has
been discussed extensively in Ref.~\onlinecite{read-rezayi}, so this is not a new observation. However, this observation
will allow us to understand the origin of the superextensive term $\frac{1}{2}m N^2$ in
$\lan\psi_0|\mathsf{\Lambda}^{ab}|\psi_0\ran$, and to explain why it should be subtracted when computing the Hall viscosity
of the CSMM. 

We start by deriving an expression for the angular momentum in the CSMM theory. To do this we use the fluid interpretation of 
the NCCS theory and CSMM from the last part of Sec.~\ref{sec:NCCS}. Our derivation of the expression for the angular
momentum consists of several steps. First, we derive an expression for the angular momentum of a classical fluid 
of charged particles on a commutative space $\mathbb{R}^2$ and in the presence of a constant background magnetic
field. Next, we take the limit in which the mass of the particles making up the fluid goes to zero. 
We then perform the noncommutative 
deformation of the expression for the angular momentum
to obtain an expression for the angular momentum in NCCS theory. Finally, the 
expression for the angular momentum in NCCS theory can also be used for the CSMM, after we replace the infinite-dimensional
operator variables in the NCCS theory with the $N\times N$ matrix variables of the CSMM.

We start with the action for a fluid of particles of mass $M$, charge $q=-e$, and constant (initial) density $\rho_0$ in a 
constant magnetic field $B$ (see the discussion in the last subsection of Sec.~\ref{sec:NCCS}), 
\beq
	S = \int_0^T dt\int d^2\mb{x}\ \rho_0 \left(\frac{1}{2}M \delta_{ab} \dot{X}^a\dot{X}^b - \frac{eB}{2}\ep_{ab} X^a\dot{X}^b \right)\ , 
\eeq
where we remind the reader that for the classical fluid the fields $X^a(t,\mb{x})$ are ordinary functions of time $t$ and
spatial coordinates $\mb{x}\in\mathbb{R}^2$. For now we omit the Lagrange multiplier field $A_0(t,\mb{x})$ which keeps
the density fixed to $\rho_0$ at all times, as this term plays no role in the definition of the angular momentum of the
theory. The momentum variables $P_{a}(t,\mb{x})$ canonically conjugate to $X^a(t,\mb{x})$ are obtained by differentiating 
the Lagrangian\footnote{We define the Lagrangian $\mathcal{L}$ by $S=\int dt \int d^2\mb{x}\ \rho_0\ \mathcal{L}$.} with 
respect to $\dot{X}^a$, and we have
\beqa
	P_1 &=& M\dot{X}^1 +\frac{eB}{2}X^2 \\
	P_2 &=& M\dot{X}^2 -\frac{eB}{2}X^1 \ .
\eeqa

The expression for the angular momentum of this fluid is then 
\beqa
	L_z &=& \int d^2\mb{x}\ \rho_0 \left( X^1 P_2 - X^2 P_1\right) \nnb \\
	 &=& \int d^2\mb{x}\ \rho_0 \left\{ M\ep_{ab}X^a\dot{X}^b - \frac{eB}{2}\delta_{ab}X^a X^b\right\}\ ,
\eeqa
and the limit $M\to 0$ gives
\beq
	L_z =  -\int d^2\mb{x}\ \rho_0 \frac{eB}{2}\delta_{ab}X^a X^b\ . \label{eq:ang-mom-M-zero}
\eeq
Next, we set $\rho_0=\frac{1}{2\pi\theta}$ as is appropriate for the fluid interpretation of NCCS theory, and 
we perform the noncommutative deformation of this expression (see Sec.~\ref{sec:NCCS}) by replacing 
$\frac{1}{2\pi\theta}\int d^2\mb{x}\ (\cdots) \to \text{Tr}_{\mathcal{H}_F}\left\{ \cdots\right\}$ and
$X^a(t,\mb{x})\to \hat{X}^a(t)$. This gives an expression for the angular momentum in NCCS theory,
\beq
	L_z= -\frac{e B}{2}\text{Tr}_{\mathcal{H}_F}\left\{ \delta_{ab}\hat{X}^a\hat{X}^b\right\}\ .
\eeq
Finally, we obtain an expression for the angular momentum of the CSMM by replacing the operators
$\hat{X}^a(t)$ with the $N\times N$ matrix variables $X^a(t)$ of the CSMM, and by replacing the trace over
the infinite-dimensional space $\mathcal{H}_F$ by the trace for $N\times N$ matrices, 
\beq
	L_{z,\text{{\tiny{CSMM}}}}= -\frac{eB}{2}\text{Tr}\{\delta_{ab}X^a X^b\}\ .
\eeq

We now compute the angular momentum in the quantum ground state $|\psi_0\ran$ of the CSMM. We first use the
expansion of Eq.~\eqref{eq:expansion} to write $L_{z,\text{{\tiny{CSMM}}}}$ as 
\beqa
	L_{z,\text{{\tiny{CSMM}}}} &=& -\frac{e B}{2}\sum_{A=0}^{N^2-1} \delta_{ab}x^a_A x^b_A \nnb \\
	&=& -\hbar \delta_{ab}\mathsf{\Lambda}^{ab}\ ,
\eeqa
where $\mathsf{\Lambda}^{ab}$ are the strain generators for the CSMM introduced in Sec.~\ref{sec:CSMM-viscosity}.
We see that our derivation of the angular momentum for the CSMM theory makes sense since 
$\delta_{ab}\mathsf{\Lambda}^{ab}$
is exactly the operator which generates rotations of the noncommutative coordinates $X^a$ in the CSMM.

For the ground state of the CSMM we have $L_{z,\text{{\tiny{CSMM}}}}|\psi_0\ran = L_0|\psi_0\ran$ with
\beq
	L_0= -\hbar\left[\frac{1}{2}m N^2 + \left( \frac{1-m}{2} \right)N \right]\ ,
\eeq
and our previous results for $\lan\psi_0|\mathsf{\Lambda}^{ab}|\psi_0\ran$ and $\eta_{\text{{\tiny{CSMM}}}}$ 
can be rewritten in the form
\beqa
	\lan\psi_0|\mathsf{\Lambda}^{ab}|\psi_0\ran &=& -\frac{1}{2\hbar}L_0\delta^{ab} \\
	\eta_{\text{{\tiny{CSMM}}}} &=& \frac{1}{2}\frac{L_0}{A}\ .
\eeqa
Thus, we see that the Hall viscosity coefficient $\eta_{\text{{\tiny{CSMM}}}}$ (before regularization) is equal to
one half the angular momentum density $\frac{L_0}{A}$ in the ground state of the CSMM (compare with the
angular momentum interpretation of the Hall viscosity from Ref.~\onlinecite{read-rezayi}). Finally, we also note
that $L_0$ is exactly the guiding center part of the angular momentum of the Laughlin $\nu=\frac{1}{m}$ state.
In the lowest Landau level the Landau orbit contribution to the angular momentum is simply $\hbar\frac{N}{2}$, which
leads to the total angular momentum of the Laughlin state 
$L_{z,\nu=\frac{1}{m}}= \hbar\left[-\frac{1}{2}mN^2+m\frac{N}{2}\right]$.

We now give a fluid interpretation of the superextensive (order $N^2$) term in $L_0$, which is equal to 
$-\frac{1}{2}\hbar m N^2$. This can be 
rewritten in terms of the density $\rho_0= \frac{1}{2\pi \ell_B^2 m}$ and radius 
$R^2_N \approx 2\ell_B^2 m N$ of the droplet described by the CSMM as
\beq
	-\frac{\pi}{4}e B\rho_0 R^4_N\ .	
\eeq
This is exactly the angular momentum of a droplet of radius $R_N$ of the classical fluid described by
the small $\theta$ limit of the NCCS action in the presence of an additional parabolic potential, as we now describe. 

Recall that in the small $\theta$ limit the NCCS theory is described by the fluid action of Eq.~\eqref{eq:NCCS-small-theta}.
Let us add to this action a parabolic potential term which is the commutative analogue of the 
potential term in the CSMM action, 
\beq
	S_{para}= -\frac{eB\tomega}{2}\rho_0 \int_0^T dt\int d^2\mb{x}\ \delta_{ab}X^a X^b\ ,
\eeq
where $\rho_0=\frac{1}{2\pi\theta}$. The equations of motion which result from Eq.~\eqref{eq:NCCS-small-theta}
plus $S_{para}$ are $\dot{X}^1=\tomega X^2$ and $\dot{X}^2= -\tomega X^1$, as well as the constant density
constraint enforced by $A_0$. For the initial condition $X^a(0,\mb{x})=x^a$ the solution to these equations
can be expressed concisely as 
\beq
	X^1(t,\mb{x})+i X^2(t,\mb{x}) = (x^1 + i x^2) e^{-i\tomega t}\ .
\eeq
Finally, using Eq.~\ref{eq:ang-mom-M-zero} for the angular momentum we find that a droplet of radius $\mathcal{R}$
has angular momentum
\beqa
	L_{orb} &=& -\int_{|\mb{x}|\leq \mathcal{R}} d^2\mb{x}\ \rho_0 \frac{eB}{2}\delta_{ab}X^a X^b \nnb \\
	&=& -\frac{\pi}{4}e B\rho_0 \mathcal{R}^4,
\eeqa
where ``orb" stands for ``orbital"  since this angular momentum is associated with an overall rotation of the
fluid. 

We see that the superextensive term in $L_0$ is exactly the orbital angular momentum of a classical fluid on a commutative
space in a magnetic field undergoing uniform rotational motion. Based on this observation, 
and using the \emph{anisospin} $\varsigma=\frac{m-1}{2}$ defined earlier, the full angular momentum in 
the ground state of the CSMM can be written as
\beq
	L_0= L_{orb} + \hbar\varsigma N\ .
\eeq
Now that we have identified the orbital contribution to the total angular momentum the remaining extensive term, 
which has a coefficient $\varsigma$, can be interpreted as a spin angular momentum for the $N$ particles in the
fluid, in keeping with the interpretations of Hall viscosity of 
Refs.~\onlinecite{haldane2009,haldane2011,park-haldane,read2009,read-rezayi}.

Now that we understand the connection between the expectation value $\lan\psi_0|\mathsf{\Lambda}^{ab}|\psi_0\ran$ 
and the
total angular momentum $L_0$ of the state $|\psi_0\ran$, we can give a fluid interpretation of the regularization scheme
for the guiding center Hall viscosity proposed in Ref.~\onlinecite{park-haldane}. Specifically, the regularization scheme
of  Ref.~\onlinecite{park-haldane} corresponds to subtracting the orbital contribution to $L_0$,
\beq
	\eta_{\text{{\tiny{CSMM}}},reg}= \frac{1}{2}\left(\frac{L_0-L_{orb}}{A}\right)= \frac{1}{2}\hbar\varsigma\rho_0\ .
\eeq
This can be justified by noting that the classical charged fluid in a constant magnetic field and on ordinary commutative
space does not exhibit a Hall viscosity\footnote{This can be seen directly by writing down the equations of motion for this
classical fluid in the Euler description (i.e., in terms of mass density and velocity fields), 
and then noting that no viscosity term is present. The Euler equations for a charged
fluid in a magnetic field and a general external potential appear, for example, in Eqns.~(46) and (47) 
of Ref.~\onlinecite{guralnik2001}.}, 
and so the Hall viscosity in the fluid described by the CSMM must only be due to the 
remaining terms in $L_0$ which do not have an interpretation in terms of the classical fluid on a commutative space.

\section{Hall viscosity in the presence of anisotropy}
\label{sec:intrinsic}

In this section we introduce a simple modification of the CSMM which incorporates a constant unimodular 
metric $g_{ab}$ (i.e., a constant metric with determinant equal to $1$). This metric parametrizes an anisotropy or
intrinsic geometry of a FQH state,  as discussed in the works of Haldane and 
collaborators~\cite{haldane2009,haldane2011,haldane-anisotropic,park-haldane}. As emphasized by 
Haldane~\cite{haldane2009,haldane2011}, introducing a unimodular metric $g_{ab}$ into the guiding center part of a
FQH state enables one to see the clear separation
of the full Hall viscosity tensor $\eta^{abcd}_{tot}$
into Landau orbit and guiding center contributions. When such a metric is used in the construction of the
guiding center part of a FQH state, the guiding center Hall viscosity tensor $\eta^{ab}_H$ is modified to be proportional to 
$g^{ab}$ (the inverse metric
of $g_{ab}$ with $g^{ab}g_{bc}=\delta^a_c$) instead of $\delta^{ab}$. In this section we show that for our modified
CSMM, the two-index Hall viscosity tensor $\eta^{ab}_{\text{{\tiny{CSMM}}}}$ is also modified to be proportional to 
$g^{ab}$. This confirms that our modification of the CSMM does indeed correspond to incorporating a nontrivial metric
$g_{ab}$ into the definition of the guiding center part of a FQH state. We also note here that the introduction of a second 
metric (in addition to the the metric of space) into the quantum Hall problem is exactly the starting point for the construction of 
the bi-metric theory of FQH states of Refs.~\onlinecite{GGB,gromov-son}.

The action for our modified CSMM takes the form
\begin{align}
	S_{CSMM} &= -\frac{eB}{2}\int_0^T dt\ \text{Tr}\Big\{ \ep_{ab}X^a D_0 X^b + 2\theta A_0 \nnb \\
&+ \tomega g_{ab}X^a X^b \Big\}  + \int_0^T \ov{\Psi}^{T}(i\dot{\Psi}+ A_0 \Psi)\ .
\end{align}
Note that the only change is the replacement of $\delta_{ab}$ with $g_{ab}$ in the quadratic potential term. This is the
only part of the action which could conceivably depend on a metric, since the time derivative term already uses the epsilon 
symbol $\ep_{ab}$ to contract indices. 
To quantize this system we make a change to a new set of variables $\tilde{X}^{\td{a}}$ 
which diagonalize the potential term but, crucially, obey the same commutation relations as the original variables. In other
words, the symplectic form on the phase space of this model takes the same form in the new variables
as in the old ones. Therefore the Poisson brackets and quantum commutation relations of the new variables will be identical
to those for the old variables.

To describe this change of variables we decompose the metric and inverse metric in terms of coframes $e^{\td{a}}_a$ and
frames $E^a_{\td{a}}$ as
\begin{subequations}
\beqa
	g_{ab} &=& e^{\td{a}}_a \delta_{\td{a}\td{b}} e^{\td{b}}_b \\
	g^{ab} &=& E^a_{\td{a}} \delta^{ab} E^b_{\td{b}}\ .
\eeqa
\end{subequations}
Note that we use new indices $\td{a},\td{b}=1,2$ for the internal indices of the frames and coframes. The frames and 
coframes satisfy the relations $E^a_{\td{a}}e^{\td{a}}_b = \delta^a_b$ and 
$E^a_{\td{a}}e^{\td{b}}_a = \delta^{\td{b}}_{\td{a}}$, which just express the fact that the matrices $e$ and $E$ (with
entries $e^{\td{a}}_a$ and $E^a_{\td{a}}$, respectively) are inverses of each other. In addition, it is possible to choose 
$\text{det}(e)=\text{det}(E)=1$. This can be seen as follows. First, note that the relation between $g_{ab}$ and 
$e^{\td{a}}_a$ can be expressed in matrix form as $g= e^T e$, where $g$ is the matrix with entries $g_{ab}$. 
This implies that $\text{det}(e)^2=\text{det}(g)=1$, so that $\text{det}(e)=\pm 1$. However, the
parametrization of $g$ in terms of $e$ is invariant under the transformation $e \to S e$ for any matrix $S\in O(2)$, i.e., any 
$S$ such that $S^T S=\mathbb{I}$. Then if for some reason we found a decomposition of $g$ with $\text{det}(e)=-1$, we
can always switch to a new parametrization with $\text{det}(e)=1$ by replacing $e$ with $Se$ for any $S\in O(2)$ with
$\text{det}(S)=-1$. Then, since $E=e^{-1}$ as matrices, we also guarantee that $\text{det}(E)=1$.

Using the frames and coframes we introduce new matrix variables $\td{X}^{\td{a}}$ as
\begin{subequations}
\beqa
	\td{X}^{\td{a}} &=& e^{\td{a}}_a X^a \\
	X^a &=& E^a_{\td{a}}\td{X}^{\td{a}}\ .
\eeqa
\end{subequations}
In terms of these variables we have
\beq
	g_{ab}X^a X^b= \delta_{\td{a}\td{b}}\td{X}^{\td{a}}\td{X}^{\td{b}}\ 
\eeq
and, crucially, 
\beqa
	\ep_{ab}X^a D_0 X^b &=& \ep_{ab}E^a_{\td{a}}E^b_{\td{b}}\td{X}^{\td{a}}D_0 \td{X}^{\td{b}} \nnb \\
	&=& \text{det}(E)\ep_{\td{a}\td{b}}\td{X}^{\td{a}}D_0 \td{X}^{\td{b}}  \nnb \\
	&=& \ep_{\td{a}\td{b}}\td{X}^{\td{a}}D_0 \td{X}^{\td{b}}\ .
\eeqa
We can then carry out the quantization of this modified CSMM using the $\tilde{X}^{\td{a}}$ variables in exactly the
same way that we quantized the original CSMM in Sec.~\ref{sec:CSMM}. For example we would start by 
expanding the $\tilde{X}^{\td{a}}$ in terms of a new set of real scalar variables $\td{x}^{\td{a}}_A$ ($A=0,\dots,N^2-1$)
exactly as in Eq.~\eqref{eq:expansion}. This procedure results in a new ground state
$|\td{\psi}_0\ran$ for the modified CSMM depending on the unimodular metric $g_{ab}$.

We can now calculate the Hall viscosity in this modified CSMM. The setup for this calculation is the
same as in Sec.~\ref{sec:CSMM-viscosity} and, in particular, we still apply an APD (or strain) to 
the physical position variables $X^a$ and 
not the new variables $\td{X}^{\td{a}}$. The final expression for the two-index Hall viscosity tensor 
$\eta^{ab}_{\text{{\tiny{CSMM}}}}$ is now proportional to the expectation value of the strain generators 
$\mathsf{\Lambda}^{ab}$ in the ground state $|\td{\psi}_0\ran$ of the modified CSMM,
\beq
	\eta^{ab}_{\text{{\tiny{CSMM}}}} = -\frac{\hbar}{A}\lan \td{\psi}_0|\mathsf{\Lambda}^{ab}|\td{\psi}_0\ran\ .
\eeq
The expectation value $\lan \td{\psi}_0|\mathsf{\Lambda}^{ab}|\td{\psi}_0\ran$ is easily computed by writing 
$\mathsf{\Lambda}^{ab}= E^a_{\td{a}}E^b_{\td{b}}\td{\mathsf{\Lambda}}^{\td{a}\td{b}}$, where
\beq
	\td{\mathsf{\Lambda}}^{\td{a}\td{b}}= \frac{1}{4\ell_B^2}\sum_{A=0}^{N^2-1}\{\td{x}^{\td{a}}_A,\td{x}^{\td{b}}_A\}
\eeq
are the strain generators for the new variables, and by noting that
\beq
	\lan \td{\psi}_0|\td{\mathsf{\Lambda}}^{\td{a}\td{b}}|\td{\psi}_0\ran= \frac{1}{2}\left[\frac{1}{2}m N^2 + \left( \frac{1-m}{2} \right)N\right]\delta^{\td{a}\td{b}}\ ,
\eeq
which follows since all quantities here are in terms of the new ``tilde" variables. Then the original expectation value of interest
evaluates to 
\beqa
	\lan \td{\psi}_0|\mathsf{\Lambda}^{ab}|\td{\psi}_0\ran &=& \frac{1}{2}\left[\frac{1}{2}m N^2 + \left( \frac{1-m}{2} \right)N\right]\delta^{\td{a}\td{b}} E^a_{\td{a}}E^b_{\td{b}} \nnb \\
	&=& \frac{1}{2}\left[\frac{1}{2}m N^2 + \left( \frac{1-m}{2} \right)N\right]g^{ab}\ .
\eeqa
After regularization, which consists of subtracting off the order $N^2$ term in this expectation value, the
Hall viscosity tensor for the modified CSMM takes the form
\beq
	\eta^{ab}_{\text{{\tiny{CSMM}}},reg}= -\frac{\hbar}{A}\frac{1}{2}\left( \frac{1-m}{2} \right)N g^{ab} = \eta_{\text{{\tiny{CSMM}}},reg} g^{ab}\ ,
\eeq
where $\eta_{\text{{\tiny{CSMM}}},reg}= \frac{1}{2}\hbar\varsigma\rho_0$ as before, and where we defined 
$\rho_0=\frac{N}{A}$. We find that the Hall viscosity tensor for the modified CSMM is exactly the guiding center
part of the Hall viscosity tensor of the Laughlin state with nontrivial guiding
center metric $g_{ab}$~\cite{haldane2011,park-haldane}.

We close this section by calculating the area $A$ and the shape of the droplet of fluid described by the ground 
state $|\td{\psi}_0\ran$ of the modified CSMM. To do this we follow the method from the end of Sec.~\ref{sec:CSMM}
and consider the eigenvalue of $\text{Tr}\left\{ g_{ab}X^a X^b\right\}$ when acting on the state $|\td{\psi}_0\ran$. We again
find that $\text{Tr}\left\{ g_{ab}X^a X^b\right\}|\td{\psi}_0\ran= R^2|\td{\psi}_0\ran$ with the same eigenvalue
$R^2$ from Eq.~\eqref{eq:droplet-radial-positions}, and we can again interpret $R^2$ as a sum of contributions from 
$N$ particles, $R^2= \sum_{j=1}^{N}R^2_j$ with $R^2_j= 2\ell_B^2\left(m (j-1) + \frac{1}{2}\right)$. However,
the interpretation of the shape of the droplet is different now since $g_{ab}X^a X^b$ is a general quadratic form
of the noncommutative position coordinates. In the simple case where $g_{ab}=\delta_{ab}$, we argued that the
droplet was circular, with the $j^{th}$ particle located somewhere on a circle of radius $R_j$. In this case we will
argue that the droplet has the shape of an ellipse, with the particular geometry of the ellipse determined by the
eigenvectors and eigenvalues of the metric $g_{ab}$ considered as a matrix, and where the $j^{th}$ particle is now
located somewhere on an ellipse whose size is determined by $R_j$.

To facilitate this analysis we use a convenient parametrization~\cite{haldane-anisotropic} of the unimodular metric $g_{ab}$ in 
terms of a single complex parameter $\gamma\in\mathbb{C}$, $|\gamma|<1$, and write
\beq
	g= \frac{1}{1-|\gamma|^2}\begin{pmatrix}
	(1+\gamma)(1+\ov{\gamma}) & i (\gamma-\ov{\gamma}) \\
	i (\gamma-\ov{\gamma})  & (1-\gamma)(1-\ov{\gamma})
\end{pmatrix}\ .
\eeq
If we also write $\gamma= \tanh(\frac{\al}{2})e^{i\beta}$ for real $\al>0$ and a real phase $\beta$, then we find that
the matrix $g$ has the decomposition 
\beq
	g= S D S^{T}
\eeq
with
\beq
	S= \begin{pmatrix}
	\cos(\frac{\beta}{2}) & \sin(\frac{\beta}{2}) \\
	-\sin(\frac{\beta}{2}) & \cos(\frac{\beta}{2})
\end{pmatrix}
\eeq
and
\beq
	D=\begin{pmatrix}
	e^{\al} & 0 \\
	0 & e^{-\al}
\end{pmatrix}\ .
\eeq
Here $e^{\pm \al}$ are the eigenvalues of $g$ and the columns of the matrix $S$ are the normalized eigenvectors of $g$.
In component form we can also write
\beq
	g_{ab}= S^{\td{a}}_a D_{\td{a}\td{b}}S^{\td{b}}_b\ ,
\eeq
where for $S^{\td{a}}_a$, $a$ indexes the rows of the matrix $S$ and $\td{a}$ indexes the columns. 

\begin{figure}[t]
  \centering
    \includegraphics[width= .3\textwidth]{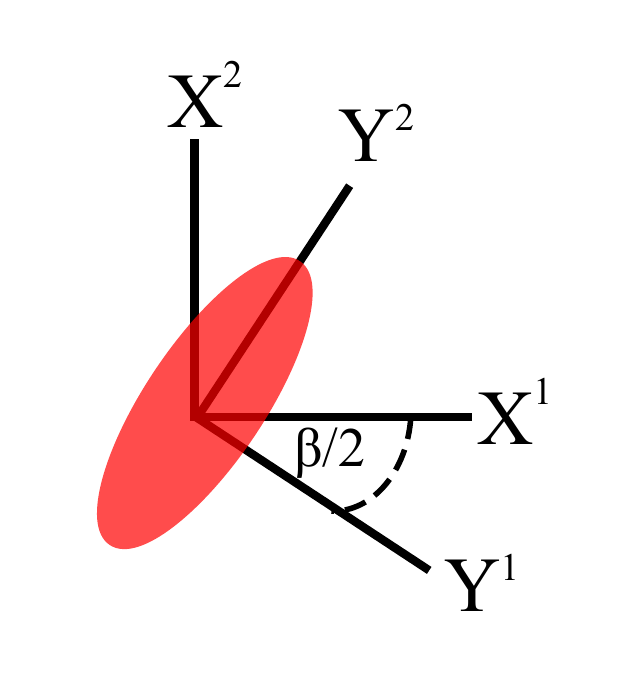} 
\vskip 10pt
 \caption{The shape and orientation of the droplet of fluid which is described by the ground state $|\td{\psi}_0\ran$
of the modified CSMM incorporating the unimodular spatial metric $g_{ab}$.}
\label{fig:ellipse}
\end{figure}

We now introduce new noncommutative coordinates (i.e., matrices) $Y^{\td{a}}$ defined as
\beq
	Y^{\td{a}}= S^{\td{a}}_a X^a\ ,
\eeq
and in terms of these we have
\beqa
	g_{ab}X^a X^b &=& D_{\td{a}\td{b}}Y^{\td{a}}Y^{\td{b}} \nnb \\
	 &=& e^{\al} (Y^1)^2 + e^{-\al}(Y^2)^2\ .
\eeqa
We now see that in the modified CSMM with metric $g_{ab}$, we can interpret the $j^{th}$ particle
as residing on an ellipse with the lengths of the minor and major axes of that ellipse given by
$r_{1,j}= e^{-\frac{\al}{2}}R_j$ and $r_{2,j}= e^{\frac{\al}{2}}R_j$\footnote{Recall that the equation 
$a^2 x^2 + b^2 y^2= R^2$ describes an ellipse in the $(x,y)$ plane with the lengths of the two axes of the ellipse given
by $\frac{R}{a}$ and $\frac{R}{b}$.}. 
Furthermore, this ellipse has its minor and major axes lined up with
the axes of the $Y^{\td{a}}$ coordinate system, which is rotated from the $X^a$ coordinate system by an angle of
$\frac{\beta}{2}$ as shown in Fig.~\ref{fig:ellipse}. The area of the ellipse where the $j^{th}$ particle is located is
$\pi r_{1,j} r_{2,j} = \pi R^2_j$, and since $R^2_j$ is linear in $j$, we again find that
the particle density is constant inside the droplet. 
Finally, the area of the droplet is equal to the area of the ellipse for particle $N$ which 
is $A = \pi R^2_N \approx 2\pi \ell_B^2 m N$, just as in the ordinary CSMM. 

We conclude that the modified CSMM
incorporating the unimodular metric $g_{ab}$ describes an elliptical droplet of fluid with the same area $A$ and constant
density $\rho_0$ as the ordinary CSMM, and where the details of the shape of the ellipse are determined by the eigenvalues
and eigenvectors of the metric $g_{ab}$. In addition, since the density $\rho_0$ is the same as for the original
CSMM, we find that the coefficient 
$\eta_{\text{{\tiny{CSMM}}},reg}= \frac{1}{2}\hbar\varsigma\rho_0$ of Hall viscosity for the
CSMM with $g_{ab}\neq \delta_{ab}$ is numerically equal to the coefficient for the case where $g_{ab}=\delta_{ab}$. The
only difference between these two cases is the structure of the Hall viscosity tensor, since for $g_{ab}\neq\delta_{ab}$ the
two index tensor $\eta^{ab}_{\text{{\tiny{CSMM}}},reg}$ is proportional to $g^{ab}$ instead of $\delta^{ab}$.
	
\section{Conclusion}
\label{sec:conclusion}

In this paper we investigated the geometric properties of the Laughlin FQH states within the CSMM description of 
these states which, roughly speaking, models these states as a charged fluid in a magnetic field and propagating on a 
noncommutative space. 
We focused our attention on the specific properties of Hall viscosity, Hall conductance in a non-uniform electric field,
and the Hall viscosity in the presence of anisotropy. We found that the answers for these quantities calculated from
the CSMM description contain only the \emph{guiding center} contribution to the known answers for these quantities in the 
Laughlin states. 

These results lead us to the general conclusion that the CSMM description of the Laughlin FQH states 
accurately captures the guiding center contribution to the geometric properties of these states, but lacks the 
Landau orbit contribution. As we remarked in the Introduction, the Landau orbit contribution is often considered to be a
trivial contribution since the interesting correlations in the Laughlin state are contained in the guiding center part of its
wave function/state vector. Therefore we find that the CSMM description captures the most important contribution, namely the 
guiding center contribution, to the physics of the Laughlin FQH states. However, any attempt to completely describe
the Laughlin states using the CSMM or NCCS theory must also include some auxiliary degrees of freedom which account for
the missing Landau orbit contributions to the geometric properties of these states.

There are several possible directions for future work in this area. One direction would be to continue to develop the
fluid interpretation of the CSMM. One goal of this work would be to find an appropriate definition of a density 
operator $\rho(\mb{x})$ which is a function of a commutative two-dimensional coordinate $\mb{x}\in\mathbb{R}^2$
and which is defined on length scales much larger than the scale set by $\theta$ in the noncommutative theory. One 
could then check whether this density operator satisfies the Girvin-Macdonald-Plaztman algebra, and also attempt to
compute the static structure factor and compare to the known answer for the Laughlin states~\cite{GMP}. 
 Another goal of this work would be to connect the CSMM description of the Laughlin states with 
a different fluid description of these states, which is Wiegmann's vortex fluid description~\cite{wiegmann2013}.
In this description the Laughlin FQH state with $N$ electrons is modeled as a rotating incompressible fluid containing $N$ point 
vortices each carrying a quantized circulation $\Gamma$ which depends on the filling fraction of the Laughlin state.
On this topic we note that Bettelheim has recently introduced a method for defining density and velocity fields in the
CSMM which are functions of a commutative coordinate $\mb{x}$ in Ref.~\onlinecite{bettelheim2013}, and it
would be interesting to develop his approach further and to use it to connect with Wiegmann's vortex fluid description. 
We also note that the problem of defining density operators in NCCS theory and the CSMM has been considered before in 
Refs.~\onlinecite{hansson2001,hansson2003}.

A second direction for future work would be to investigate the Hall viscosity and other geometric response properties
 in matrix models which
describe other more complicated FQH states. For example, a matrix model for the Jain states~\cite{jain1989} 
has been proposed in 
Ref.~\onlinecite{cappelli-jain}. More recently, the authors of Ref.~\onlinecite{tong2016} proposed a class of 
matrix models for the \emph{Blok-Wen} series of non-Abelian FQH states~\cite{blok-wen}. 
It would also be interesting to search for new matrix models which can describe
other FQH states of interest.

%
%Finally, it would be interesting to understand the exact relation between the CSMM in the $N\to \infty$ limit and the
%quantization of the full NCCS theory. It is suspected that the quantum Hilbert space of the NCCS theory has
%a single physical state (see, for example, the comments on Pg.~6 of Ref.~\onlinecite{cappelli2005}). This expectation
%is consistent with the fact that the $U(1)$ NCCS theory is mapped to the ordinary $U(1)$ Chern-Simons theory under
%the Seiberg-Witten map~\cite{grandi-silva}, and the fact that ordinary $U(1)$ Chern-Simons theory on the full plane 
%$\mathbb{R}^2$ has a unique\footnote{On a disk geometry the Hilbert space of ordinary $U(1)$ Chern-Simons
%theory has an infinite-dimensional Hilbert space corresponding to the Hilbert space of the chiral boson which lives
%at the edge of the disk. However, as shown in Ref.~\onlinecite{dunne1989}, the theory on the full plane $\mathbb{R}^2$ has 
%a unique ground state.} quantum state~\cite{dunne1989}. It would be interesting to understand how exactly the ground
%state of the CSMM is related to the (suspected) unique quantum state of the full NCCS theory. 

\acknowledgements

We acknowledge useful discussions with E. Fradkin, R. Leigh, B. Bradlyn, A. Gromov, M. Stone, T. Zhou, P. Di Francesco, 
R. Kedem, P. Wiegmann, and V. Pasquier. 
M.F.L. wishes to acknowledge the hospitality of the Simons Center for Geometry and Physics during the 2017
 workshop ``Strongly correlated topological phases of matter", 
and the hospitality of the Institut d'Etudes Scientifiques in Carg\`ese
during the 2017 ``Exact methods in low-dimensional physics" summer school, where parts of this research were conducted.
M.F.L. and T.L.H. acknowledge support from the US National
Science Foundation under grant DMR 1351895-CAR. We also gratefully acknowledge the support of the Institute
for Condensed Matter Theory at the University of Illinois at Urbana-Champaign.

\appendix

\section{Quantum generators of the $U(N)$ action}
\label{app:generators}

In this Appendix we consider the form of the quantum generators of the $U(N)$ transformations of the matrix
model variables $X^a$ and $\Psi$. We use this result in Sec.~\ref{sec:CSMM} to show that the
constraint of Eq.~\eqref{eq:constraint} simply forces physical states in the CSMM to be singlets under the $SU(N)$ action, and 
to carry a certain total charge under the $U(1)$ action. This information is sufficient to write down a basis of physical
states (states respecting the constraint) for the model following Ref.~\onlinecite{HVR}.

We start with the generators for the $U(N)$ transformation of the complex vector variable $\Psi$. Under a $U(N)$
transformation by a matrix $V$ we have $\Psi \to V\Psi$ or in components
\beq
	\Psi^j \to {V^j}_k \Psi^k\ .
\eeq
We are interested in the infinitesimal form of this transformation, so we take $V= e^{iT}$ for a Hermitian matrix $T$ (the
Lie algebra of the group $U(N)$ consists of the $N\times N$ Hermitian matrices). Then to first order in 
$T$ we have $\Psi \to \Psi + i T\Psi$. In components, the first order change in $\Psi^j$ generated by $T$ is
\beq
	\delta_{T}\Psi^j = i {T^j}_k\Psi^k\ .
\eeq

We now look for a quantum operator $\mathcal{O}_{\Psi}(T)$ such that 
\beq
	[\mathcal{O}_{\Psi}(T),\Psi^j] =  i {T^j}_k\Psi^k\ ,
\eeq
i.e., the quantum commutator of $\mathcal{O}_{\Psi}(T)$ with $\Psi^j$ implements the infinitesimal $U(N)$ action generated
by $T$ (this is what we mean when we say that a quantum operator \emph{generates} the $U(N)$ action).
The correct operator is (in terms of $b^j$ instead of $\Psi^j$)
\beq
	\mathcal{O}_{\Psi}(T)= -i b^{\dg}_j {T^j}_k b^k\ .
\eeq
Thus, $\mathcal{O}_{\Psi}(T)$ is the quantum operator which generates the $U(N)$ transformation $V=e^{iT}$ acting 
on $\Psi$. One can also check that the operators $\mathcal{O}_{\Psi}(T)$ obey the Lie algebra of $U(N)$. To check
this it is sufficient to check that the map $T\mapsto \mathcal{O}_{\Psi}(T)$ is a Lie algebra homomorphism, i.e., that
\beq
	[\mathcal{O}_{\Psi}(T_1),\mathcal{O}_{\Psi}(T_2)]= \mathcal{O}_{\Psi}(-i[T_1,T_2]_M)\ ,
\eeq
and it is straightforward to verify that this relation holds for our generators $\mathcal{O}_{\Psi}(T)$.

Next we consider the matrix variables $X^a$. Under a $U(N)$ transformation we have
$X^a\to V X^a \ov{V}^T$. Writing $V=e^{iT}$ as before, we find that to first order in $T$ we have 
$X^a\to X^a + i[T,X^a]_M$. Note that for $T= \al \mathbb{I}$, i.e., for $U(1)$ transformations, the
matrix variables $X^a$ are invariant. Therefore we can restrict our attention to $SU(N)$ transformations for the
$X^a$ variables. We then choose $T$ to be one of the generators $T^A$ of $SU(N)$, and examine the infinitesimal
action of $V= e^{iT^A}$ on the scalar variables $x^a_0$ and $x^a_A$, $A=1,\dots,N^2-1$, which appear in the expansion 
of $X^a$ from Eq.~\eqref{eq:expansion}. We have
\beqa
	\delta_{T^A} X^a &=& i[T^A,X^a]_M \nnb \\ 
	&=& i\sum_{B=1}^{N^2-1}x^a_B [T^A,T^B]_M \nnb \\
	&=& -\sum_{B,C=1}^{N^2-1} x^a_B f^{ABC} T^C\ .
\eeqa
From this we read off that $\delta_{T^A} x^a_0= 0$ (reflecting the invariance under $U(1)$ transformations), and
\beq
	\delta_{T^A} x^a_B= -\sum_{C=1}^{N^2-1} x^a_C f^{ACB}\ ,\ B=1,\dots,N^2-1\ .
\eeq

We now look for a quantum operator $\mathcal{O}_{X}(T^A)$ which generates this action on the variables $x^a_A$ 
($A=1,\dots,N^2-1$), i.e., an operator which commutes with $x^a_0$ and satisfies
\beq
	[\mathcal{O}_{X}(T^A), x^a_B]= -\sum_{C=1}^{N^2-1} x^a_C f^{ACB}
\eeq
for $B=1,\dots,N^2-1$. One can check that the correct operator is (in terms of the oscillator variables $a_A$)
\beq
	\mathcal{O}_{X}(T^A)= \sum_{B,C=1}^{N^2-1} f^{ACB} a^{\dg}_B a_C\ .
\eeq

This completes the construction of the quantum generators of the $U(N)$ action on the $X^a$ and $\Psi$ variables
in the CSMM. This is all the information which is needed to analyze the $j\neq k$ elements of the 
CSMM constraint ${G^j}_k$ from Eq.~\eqref{eq:constraint-2nd-form}.

\section{Kubo formula approach to Hall viscosity in the CSMM}
\label{app:Kubo}

In this Appendix we use a Kubo formula approach inspired by Ref.~\onlinecite{bradlyn2012} to compute the 
Hall viscosity in the ground state of the CSMM. For this computation we subject the CSMM to a 
time-dependent APD (or strain) parametrized by $\al_{ab}(t)$ such that
the dynamics of the system is described by the time-dependent Hamiltonian
\beq
	H(\al(t))= U(\al(t))H_{CSMM}U(\al(t))^{\dg}\ .
\eeq
Here the operator $U(\al(t))$ is the APD generator for the CSMM which we derive in Sec.~\ref{sec:CSMM-viscosity} 
of the main text. We also assume that at the time $t_0$ we have $\al_{ab}(t_0)= 0$ so that 
$|\psi(t_0)\ran= |\psi_0\ran$, which
is the ground state of the CSMM from Eq.~\eqref{eq:CSMM-ground-state}. As we discussed in Sec.~\ref{sec:GC-hall-visc}, 
the generalized force associated with the APD parametrized by the coefficients $\al_{ab}$ is
\beq
	F^{ab}= -\frac{\pd H(\al)}{\pd \al_{ab}}\Big|_{\al=0} = -i[\mathsf{\Lambda}^{ab},H_{CSMM}]\ .
\eeq
To calculate the Hall viscosity we need to compute the expectation value of the generalized force
$F^{ab}$ in the state $|\psi(t)\ran$ of the system, where $|\psi(t)\ran$ is the solution to the time-dependent 
Schrodinger equation
\beq
	H(\al(t))|\psi(t)\ran= i \hbar\frac{\pd}{\pd t}|\psi(t)\ran\ .
\eeq
We now discuss the details of this computation. 

First, to set up this problem in a form which is amenable to perturbation theory and the Kubo formula, we make a 
time-dependent change of states by writing
\beq
	|\psi(t)\ran= U(\al(t))|\phi(t)\ran\ .
\eeq
The state $|\phi(t)\ran$ is then the solution to a time-dependent Schrodinger equation with a new Hamiltonian $H'(t)$ given by
\beq
	H'(t)= H_{CSMM} + V(t) 
\eeq
with
\beqa
	V(t) &=& -i\hbar U(\al(t))^{\dg} \frac{\pd U(\al(t))}{\pd t} \nnb \\
	 &\approx& \hbar \frac{\pd \al_{ab}(t)}{\pd t} \mathsf{\Lambda}^{ab} + \dots\ ,
\eeqa
where in the second line we expanded the perturbation $V(t)$ to first order in $\al_{ab}(t)$. The new Hamiltonian 
$H'(t)$ is now expressed as a time-independent term $H_{CSMM}$ plus a time-dependent perturbation 
$V(t)$, and is therefore in a form\footnote{The change of basis from $|\psi(t)\ran$ to $|\phi(t)\ran$ is equivalent to the 
change from the ``$\mb{x}$" to the ``$\mb{X}$" variables in Ref.~\onlinecite{bradlyn2012}. 
We thank Barry Bradlyn for helpful discussions on this point.} 
which is amenable to an application of standard linear response theory.

To compute the Hall viscosity we naively want to compute the expectation value of $F^{ab}$ in the state
$|\psi(t)\ran$. However, in Ref.~\onlinecite{bradlyn2012} the authors argued that one should instead compute the
expectation value of $U(\al(t))F^{ab}U(\al(t))^{\dg}$, which is equivalent to expressing the generalized force $F^{ab}$ in 
terms of the strained coordinates $U(\al(t)) x^a_A U(\al(t))^{\dg}$ instead of the original coordinates $x^a_A$ of the
CSMM (in the language of Ref.~\onlinecite{bradlyn2012}, we express the generalized force in terms of the ``$\mb{X}$"
variables as opposed to the unstrained ``$\mb{x}$" variables). The reason for this is as follows. We view the
APD parametrized by $\al_{ab}(t)$ as an active transformation (i.e., we physically deform the fluid/CSMM), and so in the
computation of the response to this APD we should use the generalized force expressed in terms of the coordinates of the 
deformed system. Now we have
\beq
	\lan \psi(t) | U(\al(t))F^{ab}U(\al(t))^{\dg}|\psi(t)\ran = \lan \phi(t) |F^{ab}|\phi(t)\ran\ ,
\eeq
and so it remains to compute the expectation value $\lan \phi(t) |F^{ab}|\phi(t)\ran$.

In interaction picture perturbation theory in the strength of the potential $V(t)$, the expectation value of any 
time-independent operator $A$ in the state $|\phi(t)\ran$ is given by the standard Kubo formula as 
\begin{align}
	\lan \phi(t)|A|\phi(t)\ran-&\lan \phi(t_0)|A|\phi(t_0)\ran = \nnb \\
 -&\frac{i}{\hbar}\int_{t_0}^t dt'\ \lan \phi(t_0)|[A_I(t),V_I(t')]|\phi(t_0)\ran + \dots \ ,
\end{align}
where $A_I(t)= e^{i \frac{H_{CSMM}(t-t_0)}{\hbar}}Ae^{-i \frac{H_{CSMM}(t-t_0)}{\hbar}}$ is in the interaction picture 
defined by evolution with $H_{CSMM}$, and likewise for 
$V_I(t')= e^{i \frac{H_{CSMM}(t'-t_0)}{\hbar}}V(t')e^{-i \frac{H_{CSMM}(t'-t_0)}{\hbar}}$. Note also that for any
time-independent $A$ we have $A_I(t_0)= A$, and we also have $|\phi(t_0)\ran=|\psi(t_0)\ran= |\psi_0\ran$.

For the application to the calculation of the Hall viscosity we set $A= F^{ab}$ and keep only the term in $V(t)$ 
which is linear in the parameters $\al_{ab}(t)$. This yields the expression
\beq
	\lan F^{ab} \ran_t - \lan F^{ab}\ran_{t_0} = -i\int_{t_0}^t dt'\ \lan [F^{ab}_I(t),\mathsf{\Lambda}^{cd}_I(t')]\ran_{t_0}\frac{\pd \al_{cd}(t')}{\pd t'}\ ,
\eeq
where we used the shorthand notation $\lan F^{ab} \ran_t \equiv \lan \phi(t)|F^{ab}|\phi(t)\ran$, etc. 
Next, since $\lan [F^{ab}_I(t),\mathsf{\Lambda}^{cd}_I(t')]\ran_{t_0}= \lan [F^{ab}_I(t-t'+t_0),\mathsf{\Lambda}^{cd}_I(t_0)]\ran_{t_0}$, 
this can be rewritten as
\beq
	\lan F^{ab} \ran_t - \lan F^{ab}\ran_{t_0} = -\int_{-\infty}^{\infty} dt'\ \mathcal{X}^{abcd}(t-t')\frac{\pd \al_{cd}(t')}{\pd t'}\ , \label{eq:Kubo1}
\eeq
where we defined the response function
\beq
	\mathcal{X}^{abcd}(t)= \lim_{\ep\to 0^{+}} i \Theta(t) \lan [F^{ab}_I(t+t_0),\mathsf{\Lambda}^{cd}_I(t_0)]\ran_{t_0}e^{-\ep t}\ ,
\eeq
and where we also sent $t_0\to -\infty$ in Eq.~\eqref{eq:Kubo1}. Note that in Eq.~\eqref{eq:Kubo1} the Heaviside function 
$\Theta(t-t')$ allows us to extend the upper limit of the integral over $t'$ to $+\infty$, while the presence of the factor
$e^{-\ep(t-t')}$ allows us to send $t_0\to -\infty$. 

Next we perform a Fourier transform\footnote{Our convention for Fourier transforms is
$f(\omega)= \int_{-\infty}^{\infty} dt\ f(t) e^{i\omega t}$, 
$f(t)= \int_{-\infty}^{\infty} \frac{d\omega}{2\pi} f(\omega) e^{-i\omega t}$.} and consider the 
frequency-dependent response function
\begin{align}
	\mathcal{X}^{abcd}(\omega) &= \int_{-\infty}^{\infty}dt\  \mathcal{X}^{abcd}(t) e^{i\omega t} \nnb \\
	&= \lim_{\ep\to 0^{+}} i \int_0^{\infty} dt\ e^{i\omega_{+}t} \lan [F^{ab}_I(t+t_0),\mathsf{\Lambda}^{cd}_I(t_0)]\ran_{t_0}\ ,
\end{align}
where $\omega_{+}= \omega+i\ep$. Now we note that 
\beq
	F^{ab}_I(t+t_0)=  -i[\mathsf{\Lambda}^{ab}_I(t+t_0),H_{CSMM}]= \hbar\frac{d \mathsf{\Lambda}^{ab}_I(t+t_0)}{dt}\ ,
\eeq
where we used the equation of motion for $\mathsf{\Lambda}^{ab}_I(t+t_0)$ in the interaction picture. Then an integration by 
parts with respect to $t$ in the expression for $\mathcal{X}^{abcd}(\omega)$ yields a ``strain-strain" form of the
response function $\mathcal{X}^{abcd}(\omega)$ analogous to Eq.~(3.5) of Ref.~\onlinecite{bradlyn2012},
\begin{align}
	\mathcal{X}^{abcd}(\omega)&= -i\hbar \lan [\mathsf{\Lambda}^{ab}(t_0),\mathsf{\Lambda}^{cd}(t_0)]\ran_{t_0} \nnb \\
&+ \lim_{\ep\to 0^+}\hbar\omega_{+}\int_0^{\infty} dt\ e^{i\omega_{+}t} \lan [\mathsf{\Lambda}^{ab}(t+t_0),\mathsf{\Lambda}^{cd}(t_0)]\ran_{t_0} \ . \label{eq:strain-strain}
\end{align}
In the case where the unperturbed Hamiltonian has a unique ground state and a finite energy gap one finds that
\beqa
	\lim_{\omega\to 0}\mathcal{X}^{abcd}(\omega) &=& -i\hbar \lan [\mathsf{\Lambda}^{ab}(t_0),\mathsf{\Lambda}^{cd}(t_0)]\ran_{t_0} \nnb \\
	&=& -i\hbar \lan \psi_0| [\mathsf{\Lambda}^{ab},\mathsf{\Lambda}^{cd}]|\psi_0\ran \ ,
\eeqa
i.e., the first term in Eq.~\eqref{eq:strain-strain} gives the full response at $\omega=0$~\cite{bradlyn2012}. These
assumptions (unique ground state and finite energy gap) hold for the CSMM for any finite value of $\tilde{\omega}$, and
so this formula for the response at $\omega=0$ can be applied to the CSMM\footnote{One
should not confuse $\omega$, the frequency appearing in the Fourier transform of the response function, with $\td{\omega}$,
which sets the strength of the parabolic potential in the CSMM.}.
We note that this form
of the response at $\omega=0$ is what one obtains from a Hall viscosity calculation using adiabatic perturbation 
theory~\cite{ASZ,read2009,read-rezayi,park-haldane}.

Finally, we can complete the calculation of $\lan F^{ab} \ran_t \equiv \lan \phi(t)|F^{ab}|\phi(t)\ran$ to lowest order
in time derivatives of $\al_{cd}(t)$. First, after a Fourier transformation (taking $t_0\to -\infty$ in order to do the
integration over $t'$) we can write
\beq
	\lan F^{ab} \ran_t - \lan F^{ab}\ran_{t_0}= \int_{-\infty}^{\infty}\frac{d\omega}{2\pi}\ i\omega \mathcal{X}^{abcd}(\omega)\al_{cd}(\omega) e^{-i\omega t}\ .
\eeq
Next, we expand $\mathcal{X}^{abcd}(\omega)$ about $\omega=0$ as
\begin{align}
	\mathcal{X}^{abcd}(\omega) = -i\hbar \lan \psi_0| [\mathsf{\Lambda}^{ab},\mathsf{\Lambda}^{cd}]|\psi_0\ran + \dots
\end{align}
and invert the Fourier transformation to find
\beq
	\lan F^{ab} \ran_t - \lan F^{ab}\ran_{t_0}= i\hbar \lan \psi_0| [\mathsf{\Lambda}^{ab},\mathsf{\Lambda}^{cd}]|\psi_0\ran\frac{\pd \al_{cd}(t)}{\pd t} + \dots
\eeq
For a system with an area $A$ ($A= 2\pi\ell_B^2 m N$ for the CSMM with $\theta=\ell_B^2 m$) we then find that the Hall 
viscosity tensor is given by
\beq
	\eta^{abcd}_{\text{{\tiny{CSMM}}}}= \frac{i\hbar}{A} \lan \psi_0| [\mathsf{\Lambda}^{ab},\mathsf{\Lambda}^{cd}]|\psi_0\ran\ ,
\eeq 
and this tensor encodes the linear response of the ``generalized stress" $\frac{F^{ab}}{A}$ to the ``rate of strain" given by
$\frac{\pd\al_{cd}(t)}{\pd t}$.

%\bibliography{CSMM-refs}


\end{document}